\documentclass[aps,amsmath,amssymb, reprint, prl, preprintnumbers,superscriptaddress]{revtex4-1}

\usepackage{graphicx}%

\usepackage{hyperref} 

\usepackage{array}
\usepackage{longtable}
\usepackage{verbatim}

\newcommand{\beqa}{\begin{eqnarray}}
\newcommand{\eeqa}{\end{eqnarray}}
\newcommand{\CR}{\nonumber \\}

\newcommand{\half}{\frac{1}{2}}

\newcommand{\bea}{\begin{eqnarray}}
\newcommand{\eea}{\end{eqnarray}}

\newcommand{\CC}{{\mathcal C}}

\newcommand{\CN}{{\mathcal N}}
\newcommand{\CO}{{\mathcal O}}

\newcommand{\CS}{{\mathcal S}}
\newcommand{\CT}{{\mathcal T}}

\def\a{\alpha}

\def\CC{{\cal C}}

\def\CI{{\cal I}}

\def\CR{{\cal R}}

\def\CO{{\cal O}}

\newcommand{\be}{\begin{eqnarray}}
\newcommand{\ee}{\end{eqnarray}}

\newcommand{\bn}{\begin{enumerate}}
\newcommand{\en}{\end{enumerate}}

\def\Tr{{\rm Tr}}
\def\tr{{\rm Tr}}

	\newcommand{\ba}[1]{\begin{align} #1 \end{align} }

\begin{document}

\title{Landscape of Simple Superconformal Field Theories in 4d}

\preprint{KIAS-P18044}

\author{Kazunobu Maruyoshi}
\affiliation{Faculty of Science and Technology, Seikei University\\ 3-3-1 Kichijoji-Kitamachi, Musashino-shi, Tokyo, 180-8633, Japan}
\email{maruyoshi@st.seikei.ac.jp}

\author{Emily Nardoni}
\affiliation{Mani L. Bhaumik Institute for Theoretical Physics, Department of Physics and Astronomy, University of California, Los Angeles,  CA 90095, USA}
\affiliation{Department of Physics, University of California, San Diego,  La Jolla, CA 92093, USA}
\email{enardoni@ucla.edu}

\author{Jaewon Song}
\affiliation{School of Physics, Korea Institute for Advanced Study\\
85 Hoegiro, Dongdaemun-gu, Seoul 02455, Korea}
\email{jsong@kias.re.kr}

\begin{abstract}

We explore the space of renormalization group flows that originate from supersymmetric $\mathcal{N}=1$ $SU(2)$ gauge theory with one adjoint and a pair of fundamental chiral multiplets. By considering all possible relevant deformations---including coupling to gauge-singlet chiral multiplets---we find 34 fixed points in this simple setup.  We observe that theories in this class exhibit many novel phenomena: emergent symmetry, decoupling of operators, and narrow distribution of central charges $a/c$. This set of theories includes two of the $\mathcal{N}=2$ minimal Argyres-Douglas theories and their mass deformed versions. In addition, we find 36 candidate fixed point theories possessing unphysical fermionic operators---including one with central charges $(a, c)\simeq (0.20, 0.22)$ that are smaller than any known superconformal theory---that need further investigation.

\end{abstract}

\setcounter{tocdepth}{2}
\maketitle
\section{Introduction}
  Conformal field theory (CFT) is an important object in theoretical physics, 
  which displays the physics of the low energy fixed points of some gauge theories and of critical phenomena in condensed matter theories.
  One interesting question of CFT is to find the ``minimal" interacting theory. 
  In four dimensions, a measure of minimality is the $a$ central charge, the coefficient to the Euler density of the trace anomaly,
  due to the $a$-theorem \cite{Cardy:1988cwa,Komargodski:2011vj}, $a_{{\rm UV}}>a_{{\rm IR}}$ for all unitary renormalization group (RG) flows. 
  A related quantity is the $c$ central charge, the coefficient to the two-point function of the stress-energy tensor.
  
  However, it is difficult to analyze strongly-coupled CFTs, even $a$ and $c$, in general.
  The conformal bootstrap program \cite{Rattazzi:2008pe} partially solves this by giving a bound on $c$, 
  but does not tell what the actual minimal theory is. 
  The situation changes drastically in theories with supersymmetry.
  The superconformal symmetry allows to relate the central charges to 't Hooft anomalies of the $R$-symmetry \cite{Anselmi:1997am},
  which are determined by the $a$-maximization \cite{Intriligator:2003jj} or higher supersymmetry itself.
  
  The central charge $c$ of any unitary interacting $\CN=2$ SCFT satisfies $c\geq \frac{11}{30}$ \cite{Liendo:2015ofa}. 
  The theory that saturates the bound is the Argyres-Douglas theory \cite{Argyres:1995jj,Argyres:1995xn}, 
  denoted as $H_0$ or $(A_1,A_2)$ in the literature. 
  $H_0$ also has the smallest known value of $a$ for an interacting $\CN=2$ theory.
  
  In $\CN=1$ theories, no analytic bound on the central charges is known so far. 
  However, the numerical bootstrap suggests that the SCFT with the minimal central charge has 
  a chiral operator $\CO$ with chiral ring relation $\CO^2=0$ \cite{Poland:2011ey,Bobev:2015jxa,Poland:2015mta}, 
  and a bound $c\geq 1/9 \simeq 0.11$ \cite{Poland:2015mta}.
  Is there a theory which saturates this bound?
  The minimal theory thus far known in the literature has $a=\frac{263}{768}\simeq 0.34$ and $c=\frac{271}{768} \simeq 0.35$, 
  which was constructed via a deformation of the $H_0$ theory \cite{Xie:2016hny, Buican:2016hnq}, and thus denoted as $H_0^*$. 
  \footnote{See also a recent work on 3d $\CN=4$ theory \cite{Gang:2018huc}.}
  This value of $c$ is large compared to the bound.

  In the present work, we initiate a classification of $\CN=1$ SCFT in four dimensions obtained from  Lagrangian theories
  to find a minimal SCFT. 
  We explore the space of RG flows and fixed points that originate from the simple starting point of supersymmetric $SU(2)$ gauge theory 
  with one adjoint and a pair of fundamental chiral multiplets.
  From this minimal matter content, we consider all the possible relevant deformations, 
  including deformations by coupling gauge-singlet chiral multiplets.
  Among the fixed points we obtain, two have enhanced $\CN=2$ supersymmetry: the Argyres-Douglas theories $H_0$ and $H_1$, 
  as already found in \cite{Maruyoshi:2016tqk,Maruyoshi:2016aim,Agarwal:2016pjo}.
  The other 32 are $\CN=1$ supersymmetric, including the $H_0^*$ theory as a minimal theory in terms of $a$.
  We verify that these are ``good" theories in the sense that there is no unitary-violating operator 
  by utilizing the superconformal indices \cite{Kinney:2005ej,Romelsberger:2005eg}. 
  It is remarkable that $(a,c)$ of these SCFTs distribute within a narrow range as in Figure \ref{fig:acgood}, 
  although the allowed bound of $a/c$ is wide \cite{Hofman:2008ar}. 
  
  In addition, we find 36 candidate fixed points which have an accidental global symmetry in the infrared
  and some unphysical operators, thus we refer to them as ``bad" theories. 
  Remarkably, these include theories with even smaller central charges than those of $H_0^*$.
  The minimal one, which we denote as $\CT_M$, has  $a \simeq 0.20$, and $c \simeq 0.22$.
  Although we are not able to conclude that these bad theories are really physical by the present techniques,
  we scope their properties.

\section{A Landscape of Simple SCFTs}

We systematically enumerate a large set of superconformal fixed points via the following procedure: 

\begin{enumerate}
\item Start with some fixed point theory $\CT$. 

\item Find the set of all the relevant chiral operators of $\CT$, which we will call $\CR_{\CT} $. Let us also denote $\CS_\CT \subset \CR_\CT$ as the set of operators with $R$-charge less than $4/3$.

\item Consider the fixed points $\{ \CT_\CO \}$ obtained by the deformation $\delta W = \CO$ for all $\CO \in \CR_\CT$. 

\item Consider the fixed points $\{ {\CT}_{\overline{\CO}} \}$ given by adding an additional gauge-singlet $\CN=1$ chiral field $M$ and the superpotential coupling $\delta W=M \CO$ for all $\CO \in \CS_\CT$. 

\item For each of the new fixed point theories obtained in previous steps, check if it has an operator $\CO_d$ that decouples. Remove it by introducing an $\CN=1$ chiral field $X$ and a superpotential coupling $\delta W = X \CO_d$.
We will use this notation to clearly distinguish $X$ from $M$ in the following.
     
\item For each new fixed point, repeat the entire procedure. Terminate if there is no new fixed point. 

\end{enumerate}
We employ the $a$-maximization procedure  \cite{Intriligator:2003jj} and its modification \cite{Kutasov:2003iy} to compute the superconformal $R$-charges at each step. 
Beyond $a$-maximization, we check whether the theory passes basic tests as a viable unitary SCFT:  
one is the Hofman-Maldacena bounds for $\CN=1$ SCFTs, $\frac{1}{2}\leq \frac{a}{c}\leq \frac{3}{2}$ \cite{Hofman:2008ar};
the other one is the superconformal index.
Some of the candidate fixed points have trivial index, or violate the unitarity constraints \cite{Beem:2012yn, Evtikhiev:2017heo}. 

We perform this procedure for $SU(2)$ gauge theory with the adjoint chiral multiplet $\phi$ and two fundamental chiral multiplets, $q$ and $\tilde{q}$ ($N_f=1$). 
When there is no superpotential, this theory flows to an interacting SCFT $\hat{\CT}$, as discussed in \cite{Intriligator:1994sm} (also see \cite{Elitzur:1995xp}), and a free chiral multiplet $\Tr \phi^2$. 
To pick up only the interacting piece, we add the additional singlet $X$ and the superpotential $W_{\hat{\CT}} = X \Tr \phi^2$.

Starting from $\hat{\CT}$, we apply the deformation procedure, and find 34 non-trivial distinct fixed points. 
These theories pass every test we have checked, so we call them ``good" theories. 
One caveat is that some of these ``good" theories have a flavor symmetry that is not classically manifest.
This can be explicitly seen from the superconformal index as we will discuss shortly. In these cases we cannot rule out the possibility that this symmetry mixes with the superconformal R-symmetry causing unitarity violation.
Nevertheless we keep referring to these theories as ``good", assuming there is no such mixing. We list the theories with this feature in the appendix.

There are an additional 36 distinct theories that pass almost all of our checks, except that there is a term in the index that signals a violation of unitarity. The existence of such a term implies that either the theory does not flow to an SCFT in the IR, or the answers we obtained were incorrect because we failed to take into account an accidental symmetry. In fact, these ``bad" theories 
also have an accidental $U(1)$ symmetry which is not visible at the level of the superpotential, but is evident by the existence of the corresponding conserved current term present in the index. 
At present we do not know how to account for this accidental symmetry, and so cannot say for certain if these flows will lead to SCFTs or not.

Interestingly, 6 of these ``bad" theories appear to have central charges lower than that of $H_0^*$. Denote the lowest one $\CT_M$. This is a hint that there might be a minimal SCFT in this landscape. 

\begin{figure}[t]
\begin{center}
\includegraphics[width=3.3in]{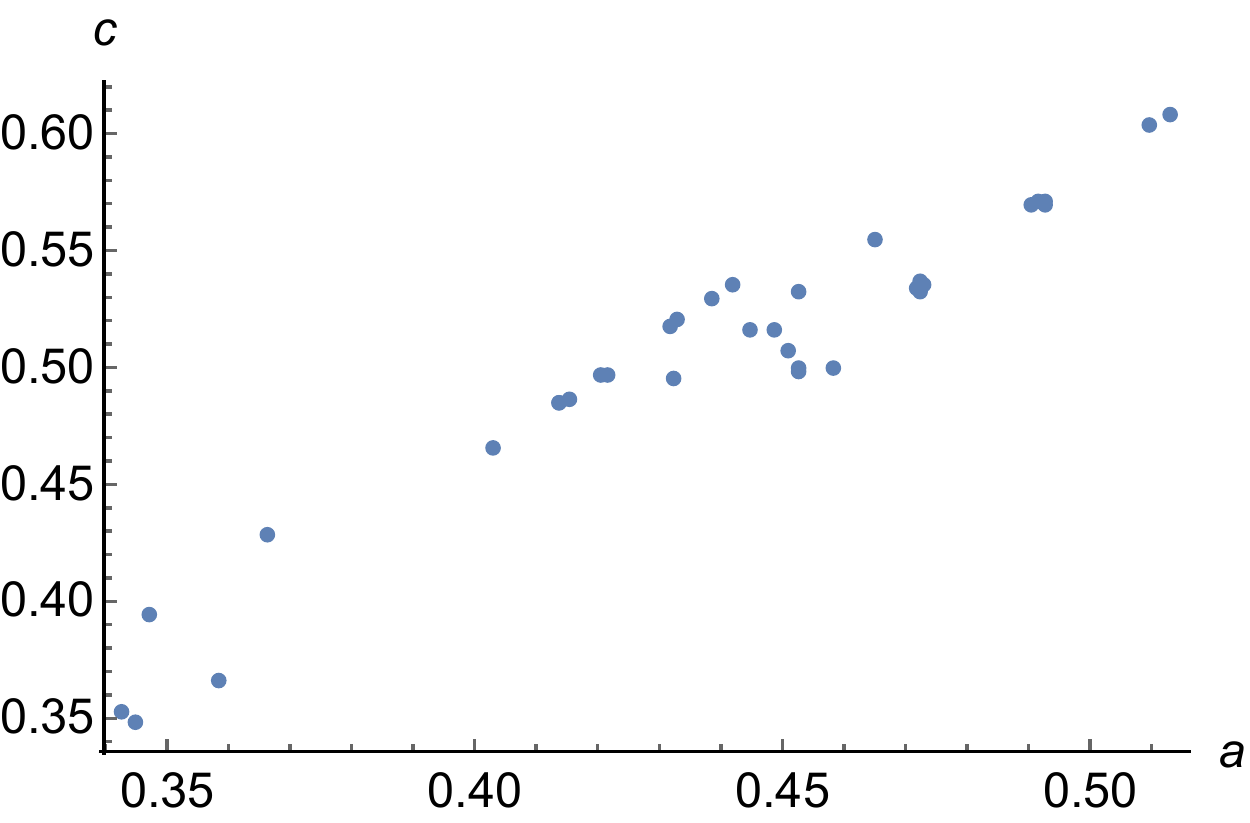}
\caption{The central charges of the 34 ``good" theories. The ratios $a/c$ all lie within the range $(0.8246,0.9895)$. The mean value of $a/c$ is $0.8732$ with standard deviation $0.0403$.}
\label{fig:acgood}
\end{center}
\end{figure}
\begin{figure}[t]
\begin{center}
\includegraphics[width=3.41in]{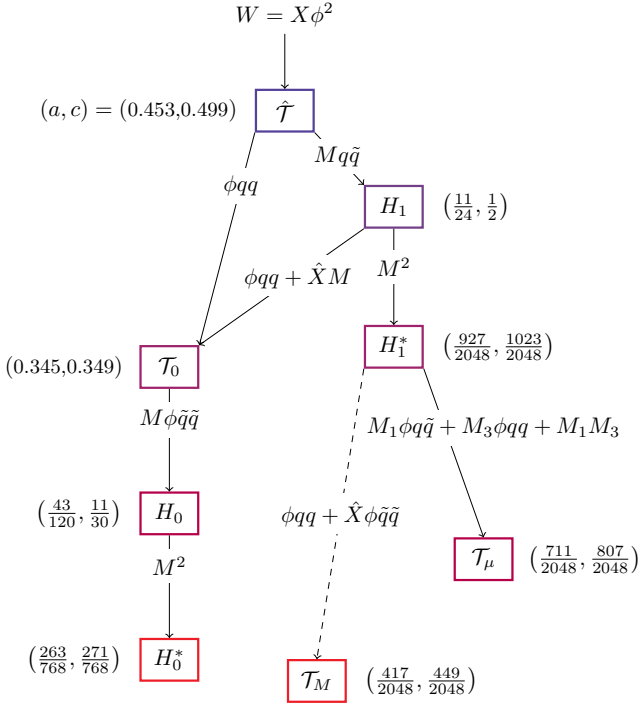}
\caption{A subset of the fixed points that can be obtained from $SU(2)$ $N_f=1$ adjoint SQCD with singlets. The arrows are labeled by the superpotential deformations. Note that the graph is not arranged vertically by decreasing $a$ central charge, because the deformations we consider involve coupling in the singlet fields. }
\label{fig:map}
\end{center}
\end{figure}

We have plotted $a,c$ for the ``good" theories without this interesting complication in Figure \ref{fig:acgood}. 
We see that the distribution of $a$ vs $c$ are concentrated near the line of $a/c \sim 0.87$.
All the theories satisfy the Hofman-Maldacena bound, and more curiously the stronger bound conjectured by \cite{Bobev:2017uzs}
$\frac{3}{5} \leq \frac{a}{c}$.
Of the ``good" theories, $H_0^*$ has the smallest value of $a$. 
$\CT_0$ has the smallest value of $a$ among any theory with a $U(1)$ flavor symmetry. 
$H_1^*$ has the smallest value of $a$ among any theory with an $SU(2)$ flavor symmetry \footnote{There are two theories with 3 conserved currents with smaller $a$, but we do not find any evidence for the $SU(2)$ symmetry.}. 
Below we examine each of these ``minimal" theories in turn, as well as the lowest central charge theory $\CT_M$, and the second-to-lowest $a$ central charge ``good" theory with no flavor symmetry, which we denote $\CT_\mu$. We summarize the structure of RG flows among these special theories in Figure \ref{fig:map}.  The full list of our theories appears in the appendix.

The superconformal indices of these theories can be computed using the Lagrangian description. 
We define the index as 
\begin{align}
\CI(t, y; x) = \Tr (-1)^F t^{3(r+2j_1)}y^{2j_2} x^f \ , 
\end{align}
where $(j_1, j_2)$ are the spins of the Lorentz group and $r$ the $U(1)$ $R$-charge. When the theory has a global symmetry with Cartan generator $f$, we also include the fugacity $x$ for it. 
For each of these special theories, we give the first few terms in the reduced superconformal index 
\begin{align}
\mathcal{I}_{r}(t,y) = (1-t^3/y)(1-t^3y)({\mathcal{I}}(t,y)-1) \ , 
\end{align} 
which removes the conformal descendant contributions coming from spacetime derivatives. 
If the reduced index contains a term $t^{R} \chi_{2j+1}(y)$ with $R<2+2j$ or a term $(-1)^{2j+1} t^R \chi_{2j+1}(y)$ with $2+2j \leq R < 6+2j$, it violates the unitarity constraint \cite{Beem:2012yn, Evtikhiev:2017heo}. 

The coefficient of $t^6 y^0 $ allows us to read off the number of marginal operators minus the number of conserved currents \cite{Beem:2012yn}. 
The superpotential F-terms $\partial W/\partial \varphi=0$ for the fields $\varphi$ allow us to read off the classical chiral ring, and quantum modifications can be argued from the index. We will see that the chiral rings we study in this paper are subject to the quantum corrections. The superconformal index turns out to be a useful tool to study the fully quantum corrected chiral rings of our models.

\section{$\CT_0$ - minimal $c$, minimal $a$ with $U(1)$}
Let us begin with the $\CT_0$ SCFT which is obtained via a deformation of $\hat{\CT}$,
	\ba{
	W_{\CT_0}=X\tr \phi^2 +\tr \phi q q,\label{eq:t0}
	} 
and has irrational central charges	
\begin{align}
\begin{split}
 	a_{\CT_0}
	&= \frac{81108+1465\sqrt{1465} }{397488}
	 \simeq 0.3451,  \\
	c_{\CT_0} 
	&= \frac{29088+1051 \sqrt{1465} }{198744}\simeq 0.3488.
\end{split}
\end{align}
The IR $R$-charges of the fields of the $\CT_0$ and all other theories discussed below are given in Table \ref{tab:Rcharges}. This theory has the second smallest value of $a$, and the smallest value of $c$ among the 34 ``good" fixed points we find \footnote{The theory with smaller $c$ than $H_0^*$ was also noticed by Sergio Benvenuti. We thank him for informing us on this.}. 
	
	\begin{table}[t]
    	\centering
    	\begin{tabular}{|c||c|c|c|c|c|}
  \hline
    \textrm{fields} & $\CT_0$ &  $H_0^*$ & $H_1^*$ & $\CT_\mu$ & $\CT_M$ \\
    \hline \hline
   $q$ & $\frac{543-\sqrt{1465}}{546} \simeq 0.924 $ & $11/12$ & $1/2$ & $1/4$  & $7/8$ \\
   \hline
   $\tilde{q}$ & $\frac{75-\sqrt{1465}}{78} \simeq 0.471$ & $5/12$ & $1/2$ & $3/4$  & $1/8$ \\
   \hline
   $\phi$ & $\frac{3+\sqrt{1465}}{273} \simeq 0.151$ & $1/6$ & $1/4$ & $1/4$  & $1/4$  \\
   \hline
   $M$ & $\cdot$ & $1$ & $1$ & $(\frac{3}{4}, 1, \frac{5}{4})$ & $1$ \\
   \hline
   $X$ & $\frac{2(270 - \sqrt{1465})}{273} \simeq 1.70$ & $5/3$ & $3/2$& $3/2$  & $3/2$ \\
   \hline
   $\hat{X}$ & $\cdot$  &$\cdot$ & $\cdot$ & $\cdot$ & $3/2$ \\
   \hline
  \end{tabular}
  \caption{The $R$-charges of the chiral multiplets at various fixed points. The $\CT_\mu$ theory has 3 chiral multiplets labeled $M$, which we denote as $M_{1, 2, 3}$.} 
    	\label{tab:Rcharges}
\end{table}

The chiral ring of the theory can be easily studied: the F-term conditions from \eqref{eq:t0} are simply 
$\tr \phi^2 = 0$, $q \phi=0$ and $X \phi + q^2 = 0$.
The first equation truncates the chiral ring by setting $\phi^2 = 0$.
The second and third equations lead to the classical generators of the chiral ring: $\CO'\equiv \tr q \tilde{q}$, $\tr \phi \tilde{q} \tilde{q}$ and $X$, with relation $\CO'^2\sim X \tr \phi \tilde{q}\tilde{q}$. 

This theory has an anomaly free $U(1)$ flavor symmetry that mixes with $R$. The reduced index is given as
\begin{align}
\begin{split}
\CI_r(t, y; x) &= t^{3.28} x^{12} - t^{3.45} x^{-2}\chi_2(y)  +t^{4.19} x^{8} -t^6 \\ 
&~~  +t^{6.56} x^{24} +t^{7.46} x^{20} + t^{8.27}x^{-10}  + \dots \ , 
\end{split}
\end{align}
where we assigned the flavor charges for the fugacity $x$ as $f_q = 1, f_{\tilde{q}}= 7, f_{\phi} = -2, f_{X} =  4$. 
Here and below $\chi_s (a)$ denotes the character for the $SU(2)$ flavor symmetry of dimension $s=2j+1$.
This index allows us to read off the quantum modified chiral ring:
the terms $t^{3.28}x^{12}$ and $t^{4.19}x^8$ in the index come from the chiral operators $\Tr\phi\tilde{q}\tilde{q}$ and $\Tr q\tilde{q}$ respectively; the second term denotes the fermionic operator $\CO_\alpha = \Tr \phi W_\alpha$. We see that the operator $X$ (which would contribute $t^{5.10} x^4$ to the index if it exists) is absent from the chiral ring \footnote{We might have anticipated this from the fact that $X$ was introduced to decouple $\tr\phi^2$, but as we see in later examples $X$ might remain a generator.}. We can read off the chiral ring relation $\CO'^2 = \CO_\alpha \cdot (\Tr \phi \tilde{q} \tilde{q}) = 0$ from the absence of the terms $t^{8.38}x^{16}$ and $-t^{6.73} \chi_2(y) x^{10}$.

\section{$H_0^*$ - minimal $a$}
The $H_0$ fixed point can be obtained from $\CT_0$ by adding the $M \Tr \phi \tilde{q} \tilde{q}$ term.
This superpotential is indeed a simplified version of the one considered in \cite{Maruyoshi:2016tqk}.
At the $H_0$ fixed point we further deform by a mass term $M^2$,
\begin{align}
W_{H_0^*}=X\tr \phi^2 + \tr \phi qq + M \tr \phi \tilde{q}\tilde{q} + M^2.\label{eq:supH0s}
\end{align}
This flows to the $H_0^*$ theory with the central charges
\begin{align}
   a_{H_0^*}=\frac{263}{768}\simeq 0.3424, \quad
   c_{H_0^*} = \frac{271}{768} \simeq 0.3529.
\end{align} 
The $H_0^*$ SCFT has been studied in \cite{Xie:2016hny, Buican:2016hnq} as a deformation of the $H_0$ Argyres-Douglas theory. Utilizing the UV Lagrangian description presented here, we are able to confirm various predictions about $H_0^*$.

Classically, the F-terms of \eqref{eq:supH0s} imply that $M,X$, and $\CO'\equiv \tr q\tilde{q}$ generate the chiral ring, with relations $M^2\sim 0$ and $\CO'^2\sim 0$. The superconformal index for the $H_0^*$ theory can be computed  to give a reduced index
\begin{align}
 \begin{split}
 \CI_r(t, y) &= t^3 - t^{\frac{7}{2}} \chi_2(y)+t^4 + t^7 +t^{\frac{17}{2}} +\dots
 \end{split}
\end{align} 
From this we see that the two generators $M$ and $\CO$ contribute the $t^3$ and $t^4$ respectively, while $X$ is not a generator.  We also find that the operator $\CO_\a = \Tr (\phi W_\a)$ contributes to $t^{\frac{7}{2}} \chi_2(y)$. From the coefficients of $t^6,t^7,t^8$, we find $M^2 = M \CO' = \CO'^2 = 0$ in the chiral ring. The term $t^7$ comes from $(\CO_\a)^2$. There is a relation for $\CO_\a$ of the form $M \CO_\a = \CO' \CO_\a = 0$ which can be read from the absence of the terms $-t^{\frac{13}{2}} \chi_2(y)$ and $-t^{\frac{15}{2}} \chi_2(y)$. These relations support the analysis of \cite{Xie:2016hny, Buican:2016hnq}. 

\section{$H_1^*$  - minimal $a$ with $SU(2)$}
The flow to $H_1$ in our setup is a simplified version of the flow considered in  \cite{Maruyoshi:2016aim}, and was also considered in \cite{Benvenuti:2017lle}. From $H_1$ the $H_1^*$ SCFT is then obtained via a mass deformation to the singlet,
\begin{align}
    W_{H_1^*}  =  X \tr \phi^2 + M \tr q \tilde{q} +   M^2.
     \label{superpotM2}
\end{align}
 The central charges are
\begin{align}
    a_{H_1^*} = \frac{927}{2048} \simeq 0.4526, \quad
    c _{H_1^*} = \frac{1023}{2048} \simeq 0.4995.
    \label{AD31charges}
\end{align}
      Classically, the F-terms imply that the chiral ring is generated by $M$, $X$, $\CO_2\equiv \tr \phi qq$, $\CO_0\equiv \tr\phi q\tilde{q}$, $\CO_{-2}\equiv\tr\phi \tilde{q}\tilde{q}$, with relations $M^2=M \CO_i = X \CO_i =0$, and $\CO_0^2\sim \CO_2\CO_{-2}$. The last relation descends from that of the Higgs branch of the $H_1$ theory. 
      
      The reduced index is
\begin{align}
\begin{split}
\CI_r(t, y; a) &=  t^3 + t^{\frac{15}{4}}(\chi_3(a)-\chi_2(y) ) + t^{\frac{9}{2}}  \\
 &~~ -  t^6 \chi_3(a) + t^{\frac{15}{2}}(1+\chi_5(a) )+ t^{\frac{33}{4}}   + \dots
\end{split}
\end{align}      
We see the theory has the $SU(2)$ current from the $- t^6 \chi_3(a) $ term, which is visible at the level of the superpotential. 
There are generators $M$, $X$ and $\CO_i$ satisfying the relations $M^2 = X^2 = 0$ and $\CO_0^2\sim \CO_2\CO_{-2}$.
There are also fermionic operators $\CO_\a = \Tr (\phi W_\a)$ with relations $M \CO_\a = X \CO_\a = 0$.

\section{$\CT_\mu$  - next to minimal}
The $\CT_\mu$ SCFT is obtained by the superpotential
\begin{align}
\begin{split}
 W &= X\Tr\phi^2 + M_2 \Tr q \tilde{q} + M_2^2 \\
  & ~~~ + M_1 \Tr \phi q \tilde{q} + M_3 \Tr \phi qq + M_1 M_3 \ ,
\end{split}
\end{align}
and the central charges are given by
\begin{align}
 a_{\CT_\mu} = \frac{711}{2048} \simeq 0.3472 , \quad c_{\CT_\mu} = \frac{807}{2048} \simeq 0.3940.  
\end{align}
The value of $a$ is the third smallest value among the ``good" theories we find and the second among the ones without flavor symmetry. 
The reduced index is
\begin{align}
 \begin{split}
 \CI_r(t, y) &= t^{\frac{9}{4}} +t^3+ t^{\frac{15}{4}}(1-\chi_2(y)) +t^{\frac{9}{2}}+t^{\frac{21}{4}} \\
 &~~ +t^{\frac{15}{2}} +  t^{\frac{33}{4}} \chi_2(y) - t^9 \chi_3(y) + \ldots 
 \end{split}
\end{align}
We see the chiral ring relations $(M_1)^3 = (M_2)^2 = (M_3)^2 = 0$. There is also an operator $\CO_\a = \Tr (\phi W_\a)$ with a relation $M_{2, 3} \CO_\a = 0$. There is no flavor symmetry as we do not have the $-t^6$ term nor a marginal operator.

\section{$\CT_M$  -- a new minimal theory?}

Let us discuss one example among the $30$ ``bad" candidate fixed points. 
Consider the superpotential
\begin{align}
 W = X \tr \phi^2 + M \tr {q \tilde{q}} + M^2 + \tr \phi q q  + \hat{X} \tr \phi \tilde{q}\tilde{q}.
\end{align}
There is no anomaly-free flavor symmetry in the Lagrangian. Assuming the $R$-charges are fixed by $W$ and anomaly condition, we get the central charges as
\begin{align}
 a_{\CT_M} = \frac{417}{2048} \simeq 0.2036, \quad c_{\CT_M} = \frac{449}{2048} \simeq  0.2192 ,  
\end{align}
and also the superconformal index of this theory as
\begin{align}
\begin{split}
\CI_r(t, y; a) &= t^3 -t^{\frac{15}{4}} \chi_2(y)+ t^{\frac{9}{2}} +t^{\frac{21}{4}}\chi_2(y) \\
 &~~- t^6 +t^{\frac{15}{2}}  - t^{\frac{33}{4}} \chi_2(y) + \dots \ .
\end{split}
\end{align}
The term $-t^6$ implies that there is a conserved current at the fixed point if it exists. The term $t^{\frac{21}{4}}\chi_2(y)$ violates the unitarity bound. As far as we know, this phenomenon has not been discussed in the literature. One possibility is that this term comes from the short multiplet $\bar{\CC}_{-\frac{1}{4}(0, \half)}$ (in the notation of \cite{Dolan:2002zh, Gadde:2010en, Beem:2012yn}) that becomes free and gets decoupled along the RG flow. Subtracting the contribution to the central charges by treating the bottom component as a free fermion with $R=-1/4$, we get the central charges $(a, c) = (\frac{189}{1024}, \frac{189}{1024}) \simeq(0.1846, 0.1846)$. We do not know if this prescription yields the correct central charges nor index of the IR theory. 

Even though we do not have a valid index, if we take it literally we can read off the chiral ring relations 
$M^2=0$, $M \widetilde{X} = 0$, where $\widetilde{X}$ is some combination of $X$ and $\hat{X}$. The other component is gone from the chiral ring.

\section{Discussion}

One goal of this program is to search for and study minimal $\CN=1$ SCFTs. 
One feature of the low-central charge SCFTs we have examined here is that there is a chiral operator satisfying a relation of the form $\CO^n\sim 0$ for $n=2,3$. 
Another feature is that the central charges of the SCFTs considered here lie in a narrow range of $a/c$.
It would be interesting to pursue the reasons for this, and search for other $\CN=1$ SCFTs with truncated chiral rings.

A common property of the RG flows in this landscape is that some operators that are irrelevant at high-energy can be relevant in the IR---such operators are called dangerously irrelevant.  
As such this is an interesting arena for studying RG flows along the lines of \cite{Gukov:2015qea}.

At present, the status of the 36 ``bad" theories is unclear, because it is not clear how to account for the accidental symmetry in the $a$-maximization procedure and thus check if the corrected theory would flow to an interacting SCFT. One way forward would be to identify the fermionic multiplet that contributes to the unitary-violating terms in the index and decouple it, as we naively did for the $\CT_M$ theory. It would be interesting to resolve this question and understand how the accidental symmetry arises.
This would settle whether one of these theories is indeed a new candidate minimal $\CN=1$ theory, or strengthen the case for minimality of the $H_0^*$ theory.

\vspace{2pt}
\begin{acknowledgments}
We would like to thank Prarit Agarwal, Matthew Buican, Sergio Benvenuti, Stefano Cremonesi, Thomas Dumitrescu, Ken Intriligator, Kimyeong Lee, Sungjay Lee,  Aneesh Manohar, Yu Nakayama, Yuji Tachikawa, and Piljin Yi for helpful discussions. JS would like to thank the UCSD High Energy Theory group for hospitality. The work of KM is supported by JSPS KAKENHI Grant Number JP17K14296. The work of EN was supported in part by DOE grant DE-SC0009919, and a UC President's Dissertation Year Fellowship. The work of JS is supported in part by  Overseas Research Program for Young Scientists through Korea Institute for Advanced Study (KIAS). Part of this work was performed at the Aspen Center for Physics, which is supported by National Science Foundation grant PHY-1607611.
\end{acknowledgments}

\bibliography{MNSbib.bib}

\begin{thebibliography}{34}%
\makeatletter
\providecommand \@ifxundefined [1]{%
 \@ifx{#1\undefined}
}%
\providecommand \@ifnum [1]{%
 \ifnum #1\expandafter \@firstoftwo
 \else \expandafter \@secondoftwo
 \fi
}%
\providecommand \@ifx [1]{%
 \ifx #1\expandafter \@firstoftwo
 \else \expandafter \@secondoftwo
 \fi
}%
\providecommand \natexlab [1]{#1}%
\providecommand \enquote  [1]{``#1''}%
\providecommand \bibnamefont  [1]{#1}%
\providecommand \bibfnamefont [1]{#1}%
\providecommand \citenamefont [1]{#1}%
\providecommand \href@noop [0]{\@secondoftwo}%
\providecommand \href [0]{\begingroup \@sanitize@url \@href}%
\providecommand \@href[1]{\@@startlink{#1}\@@href}%
\providecommand \@@href[1]{\endgroup#1\@@endlink}%
\providecommand \@sanitize@url [0]{\catcode `\\12\catcode `\$12\catcode
  `\&12\catcode `\#12\catcode `\^12\catcode `\_12\catcode `\%12\relax}%
\providecommand \@@startlink[1]{}%
\providecommand \@@endlink[0]{}%
\providecommand \url  [0]{\begingroup\@sanitize@url \@url }%
\providecommand \@url [1]{\endgroup\@href {#1}{\urlprefix }}%
\providecommand \urlprefix  [0]{URL }%
\providecommand \Eprint [0]{\href }%
\providecommand \doibase [0]{http://dx.doi.org/}%
\providecommand \selectlanguage [0]{\@gobble}%
\providecommand \bibinfo  [0]{\@secondoftwo}%
\providecommand \bibfield  [0]{\@secondoftwo}%
\providecommand \translation [1]{[#1]}%
\providecommand \BibitemOpen [0]{}%
\providecommand \bibitemStop [0]{}%
\providecommand \bibitemNoStop [0]{.\EOS\space}%
\providecommand \EOS [0]{\spacefactor3000\relax}%
\providecommand \BibitemShut  [1]{\csname bibitem#1\endcsname}%
\let\auto@bib@innerbib\@empty
\bibitem [{\citenamefont {Cardy}(1988)}]{Cardy:1988cwa}%
  \BibitemOpen
  \bibfield  {author} {\bibinfo {author} {\bibfnamefont {J.~L.}\ \bibnamefont
  {Cardy}},\ }\href {\doibase 10.1016/0370-2693(88)90054-8} {\bibfield
  {journal} {\bibinfo  {journal} {Phys. Lett.}\ }\textbf {\bibinfo {volume}
  {B215}},\ \bibinfo {pages} {749} (\bibinfo {year} {1988})}\BibitemShut
  {NoStop}%
\bibitem [{\citenamefont {Komargodski}\ and\ \citenamefont
  {Schwimmer}(2011)}]{Komargodski:2011vj}%
  \BibitemOpen
  \bibfield  {author} {\bibinfo {author} {\bibfnamefont {Z.}~\bibnamefont
  {Komargodski}}\ and\ \bibinfo {author} {\bibfnamefont {A.}~\bibnamefont
  {Schwimmer}},\ }\href {\doibase 10.1007/JHEP12(2011)099} {\bibfield
  {journal} {\bibinfo  {journal} {JHEP}\ }\textbf {\bibinfo {volume} {12}},\
  \bibinfo {pages} {099} (\bibinfo {year} {2011})},\ \Eprint
  {http://arxiv.org/abs/1107.3987} {arXiv:1107.3987 [hep-th]} \BibitemShut
  {NoStop}%
\bibitem [{\citenamefont {Rattazzi}\ \emph {et~al.}(2008)\citenamefont
  {Rattazzi}, \citenamefont {Rychkov}, \citenamefont {Tonni},\ and\
  \citenamefont {Vichi}}]{Rattazzi:2008pe}%
  \BibitemOpen
  \bibfield  {author} {\bibinfo {author} {\bibfnamefont {R.}~\bibnamefont
  {Rattazzi}}, \bibinfo {author} {\bibfnamefont {V.~S.}\ \bibnamefont
  {Rychkov}}, \bibinfo {author} {\bibfnamefont {E.}~\bibnamefont {Tonni}}, \
  and\ \bibinfo {author} {\bibfnamefont {A.}~\bibnamefont {Vichi}},\ }\href
  {\doibase 10.1088/1126-6708/2008/12/031} {\bibfield  {journal} {\bibinfo
  {journal} {JHEP}\ }\textbf {\bibinfo {volume} {12}},\ \bibinfo {pages} {031}
  (\bibinfo {year} {2008})},\ \Eprint {http://arxiv.org/abs/0807.0004}
  {arXiv:0807.0004 [hep-th]} \BibitemShut {NoStop}%
\bibitem [{\citenamefont {Anselmi}\ \emph {et~al.}(1998)\citenamefont
  {Anselmi}, \citenamefont {Freedman}, \citenamefont {Grisaru},\ and\
  \citenamefont {Johansen}}]{Anselmi:1997am}%
  \BibitemOpen
  \bibfield  {author} {\bibinfo {author} {\bibfnamefont {D.}~\bibnamefont
  {Anselmi}}, \bibinfo {author} {\bibfnamefont {D.~Z.}\ \bibnamefont
  {Freedman}}, \bibinfo {author} {\bibfnamefont {M.~T.}\ \bibnamefont
  {Grisaru}}, \ and\ \bibinfo {author} {\bibfnamefont {A.~A.}\ \bibnamefont
  {Johansen}},\ }\href {\doibase 10.1016/S0550-3213(98)00278-8} {\bibfield
  {journal} {\bibinfo  {journal} {Nucl. Phys.}\ }\textbf {\bibinfo {volume}
  {B526}},\ \bibinfo {pages} {543} (\bibinfo {year} {1998})},\ \Eprint
  {http://arxiv.org/abs/hep-th/9708042} {arXiv:hep-th/9708042 [hep-th]}
  \BibitemShut {NoStop}%
\bibitem [{\citenamefont {Intriligator}\ and\ \citenamefont
  {Wecht}(2003)}]{Intriligator:2003jj}%
  \BibitemOpen
  \bibfield  {author} {\bibinfo {author} {\bibfnamefont {K.~A.}\ \bibnamefont
  {Intriligator}}\ and\ \bibinfo {author} {\bibfnamefont {B.}~\bibnamefont
  {Wecht}},\ }\href {\doibase 10.1016/S0550-3213(03)00459-0} {\bibfield
  {journal} {\bibinfo  {journal} {Nucl. Phys.}\ }\textbf {\bibinfo {volume}
  {B667}},\ \bibinfo {pages} {183} (\bibinfo {year} {2003})},\ \Eprint
  {http://arxiv.org/abs/hep-th/0304128} {arXiv:hep-th/0304128 [hep-th]}
  \BibitemShut {NoStop}%
\bibitem [{\citenamefont {Liendo}\ \emph {et~al.}(2016)\citenamefont {Liendo},
  \citenamefont {Ramirez},\ and\ \citenamefont {Seo}}]{Liendo:2015ofa}%
  \BibitemOpen
  \bibfield  {author} {\bibinfo {author} {\bibfnamefont {P.}~\bibnamefont
  {Liendo}}, \bibinfo {author} {\bibfnamefont {I.}~\bibnamefont {Ramirez}}, \
  and\ \bibinfo {author} {\bibfnamefont {J.}~\bibnamefont {Seo}},\ }\href
  {\doibase 10.1007/JHEP02(2016)019} {\bibfield  {journal} {\bibinfo  {journal}
  {JHEP}\ }\textbf {\bibinfo {volume} {02}},\ \bibinfo {pages} {019} (\bibinfo
  {year} {2016})},\ \Eprint {http://arxiv.org/abs/1509.00033} {arXiv:1509.00033
  [hep-th]} \BibitemShut {NoStop}%
\bibitem [{\citenamefont {Argyres}\ and\ \citenamefont
  {Douglas}(1995)}]{Argyres:1995jj}%
  \BibitemOpen
  \bibfield  {author} {\bibinfo {author} {\bibfnamefont {P.~C.}\ \bibnamefont
  {Argyres}}\ and\ \bibinfo {author} {\bibfnamefont {M.~R.}\ \bibnamefont
  {Douglas}},\ }\href {\doibase 10.1016/0550-3213(95)00281-V} {\bibfield
  {journal} {\bibinfo  {journal} {Nucl. Phys.}\ }\textbf {\bibinfo {volume}
  {B448}},\ \bibinfo {pages} {93} (\bibinfo {year} {1995})},\ \Eprint
  {http://arxiv.org/abs/hep-th/9505062} {arXiv:hep-th/9505062 [hep-th]}
  \BibitemShut {NoStop}%
\bibitem [{\citenamefont {Argyres}\ \emph {et~al.}(1996)\citenamefont
  {Argyres}, \citenamefont {Plesser}, \citenamefont {Seiberg},\ and\
  \citenamefont {Witten}}]{Argyres:1995xn}%
  \BibitemOpen
  \bibfield  {author} {\bibinfo {author} {\bibfnamefont {P.~C.}\ \bibnamefont
  {Argyres}}, \bibinfo {author} {\bibfnamefont {M.~R.}\ \bibnamefont
  {Plesser}}, \bibinfo {author} {\bibfnamefont {N.}~\bibnamefont {Seiberg}}, \
  and\ \bibinfo {author} {\bibfnamefont {E.}~\bibnamefont {Witten}},\ }\href
  {\doibase 10.1016/0550-3213(95)00671-0} {\bibfield  {journal} {\bibinfo
  {journal} {Nucl. Phys.}\ }\textbf {\bibinfo {volume} {B461}},\ \bibinfo
  {pages} {71} (\bibinfo {year} {1996})},\ \Eprint
  {http://arxiv.org/abs/hep-th/9511154} {arXiv:hep-th/9511154 [hep-th]}
  \BibitemShut {NoStop}%
\bibitem [{\citenamefont {Poland}\ \emph {et~al.}(2012)\citenamefont {Poland},
  \citenamefont {Simmons-Duffin},\ and\ \citenamefont {Vichi}}]{Poland:2011ey}%
  \BibitemOpen
  \bibfield  {author} {\bibinfo {author} {\bibfnamefont {D.}~\bibnamefont
  {Poland}}, \bibinfo {author} {\bibfnamefont {D.}~\bibnamefont
  {Simmons-Duffin}}, \ and\ \bibinfo {author} {\bibfnamefont {A.}~\bibnamefont
  {Vichi}},\ }\href {\doibase 10.1007/JHEP05(2012)110} {\bibfield  {journal}
  {\bibinfo  {journal} {JHEP}\ }\textbf {\bibinfo {volume} {05}},\ \bibinfo
  {pages} {110} (\bibinfo {year} {2012})},\ \Eprint
  {http://arxiv.org/abs/1109.5176} {arXiv:1109.5176 [hep-th]} \BibitemShut
  {NoStop}%
\bibitem [{\citenamefont {Bobev}\ \emph {et~al.}(2015)\citenamefont {Bobev},
  \citenamefont {El-Showk}, \citenamefont {Mazac},\ and\ \citenamefont
  {Paulos}}]{Bobev:2015jxa}%
  \BibitemOpen
  \bibfield  {author} {\bibinfo {author} {\bibfnamefont {N.}~\bibnamefont
  {Bobev}}, \bibinfo {author} {\bibfnamefont {S.}~\bibnamefont {El-Showk}},
  \bibinfo {author} {\bibfnamefont {D.}~\bibnamefont {Mazac}}, \ and\ \bibinfo
  {author} {\bibfnamefont {M.~F.}\ \bibnamefont {Paulos}},\ }\href {\doibase
  10.1007/JHEP08(2015)142} {\bibfield  {journal} {\bibinfo  {journal} {JHEP}\
  }\textbf {\bibinfo {volume} {08}},\ \bibinfo {pages} {142} (\bibinfo {year}
  {2015})},\ \Eprint {http://arxiv.org/abs/1503.02081} {arXiv:1503.02081
  [hep-th]} \BibitemShut {NoStop}%
\bibitem [{\citenamefont {Poland}\ and\ \citenamefont
  {Stergiou}(2015)}]{Poland:2015mta}%
  \BibitemOpen
  \bibfield  {author} {\bibinfo {author} {\bibfnamefont {D.}~\bibnamefont
  {Poland}}\ and\ \bibinfo {author} {\bibfnamefont {A.}~\bibnamefont
  {Stergiou}},\ }\href {\doibase 10.1007/JHEP12(2015)121} {\bibfield  {journal}
  {\bibinfo  {journal} {JHEP}\ }\textbf {\bibinfo {volume} {12}},\ \bibinfo
  {pages} {121} (\bibinfo {year} {2015})},\ \Eprint
  {http://arxiv.org/abs/1509.06368} {arXiv:1509.06368 [hep-th]} \BibitemShut
  {NoStop}%
\bibitem [{\citenamefont {Xie}\ and\ \citenamefont
  {Yonekura}(2016)}]{Xie:2016hny}%
  \BibitemOpen
  \bibfield  {author} {\bibinfo {author} {\bibfnamefont {D.}~\bibnamefont
  {Xie}}\ and\ \bibinfo {author} {\bibfnamefont {K.}~\bibnamefont {Yonekura}},\
  }\href {\doibase 10.1103/PhysRevLett.117.011604} {\bibfield  {journal}
  {\bibinfo  {journal} {Phys. Rev. Lett.}\ }\textbf {\bibinfo {volume} {117}},\
  \bibinfo {pages} {011604} (\bibinfo {year} {2016})},\ \Eprint
  {http://arxiv.org/abs/1602.04817} {arXiv:1602.04817 [hep-th]} \BibitemShut
  {NoStop}%
\bibitem [{\citenamefont {Buican}\ and\ \citenamefont
  {Nishinaka}(2016)}]{Buican:2016hnq}%
  \BibitemOpen
  \bibfield  {author} {\bibinfo {author} {\bibfnamefont {M.}~\bibnamefont
  {Buican}}\ and\ \bibinfo {author} {\bibfnamefont {T.}~\bibnamefont
  {Nishinaka}},\ }\href {\doibase 10.1103/PhysRevD.94.125002} {\bibfield
  {journal} {\bibinfo  {journal} {Phys. Rev.}\ }\textbf {\bibinfo {volume}
  {D94}},\ \bibinfo {pages} {125002} (\bibinfo {year} {2016})},\ \Eprint
  {http://arxiv.org/abs/1602.05545} {arXiv:1602.05545 [hep-th]} \BibitemShut
  {NoStop}%
\bibitem [{Note1()}]{Note1}%
  \BibitemOpen
  \bibinfo {note} {See also a recent work on 3d ${\protect \mathcal N}=4$
  theory \cite {Gang:2018huc}.}\BibitemShut {Stop}%
\bibitem [{\citenamefont {Maruyoshi}\ and\ \citenamefont
  {Song}(2017{\natexlab{a}})}]{Maruyoshi:2016tqk}%
  \BibitemOpen
  \bibfield  {author} {\bibinfo {author} {\bibfnamefont {K.}~\bibnamefont
  {Maruyoshi}}\ and\ \bibinfo {author} {\bibfnamefont {J.}~\bibnamefont
  {Song}},\ }\href {\doibase 10.1103/PhysRevLett.118.151602} {\bibfield
  {journal} {\bibinfo  {journal} {Phys. Rev. Lett.}\ }\textbf {\bibinfo
  {volume} {118}},\ \bibinfo {pages} {151602} (\bibinfo {year}
  {2017}{\natexlab{a}})},\ \Eprint {http://arxiv.org/abs/1606.05632}
  {arXiv:1606.05632 [hep-th]} \BibitemShut {NoStop}%
\bibitem [{\citenamefont {Maruyoshi}\ and\ \citenamefont
  {Song}(2017{\natexlab{b}})}]{Maruyoshi:2016aim}%
  \BibitemOpen
  \bibfield  {author} {\bibinfo {author} {\bibfnamefont {K.}~\bibnamefont
  {Maruyoshi}}\ and\ \bibinfo {author} {\bibfnamefont {J.}~\bibnamefont
  {Song}},\ }\href {\doibase 10.1007/JHEP02(2017)075} {\bibfield  {journal}
  {\bibinfo  {journal} {JHEP}\ }\textbf {\bibinfo {volume} {02}},\ \bibinfo
  {pages} {075} (\bibinfo {year} {2017}{\natexlab{b}})},\ \Eprint
  {http://arxiv.org/abs/1607.04281} {arXiv:1607.04281 [hep-th]} \BibitemShut
  {NoStop}%
\bibitem [{\citenamefont {Agarwal}\ \emph {et~al.}(2016)\citenamefont
  {Agarwal}, \citenamefont {Maruyoshi},\ and\ \citenamefont
  {Song}}]{Agarwal:2016pjo}%
  \BibitemOpen
  \bibfield  {author} {\bibinfo {author} {\bibfnamefont {P.}~\bibnamefont
  {Agarwal}}, \bibinfo {author} {\bibfnamefont {K.}~\bibnamefont {Maruyoshi}},
  \ and\ \bibinfo {author} {\bibfnamefont {J.}~\bibnamefont {Song}},\ }\href
  {\doibase 10.1007/JHEP12(2016)103} {\bibfield  {journal} {\bibinfo  {journal}
  {JHEP}\ }\textbf {\bibinfo {volume} {12}},\ \bibinfo {pages} {103} (\bibinfo
  {year} {2016})},\ \Eprint {http://arxiv.org/abs/1610.05311} {arXiv:1610.05311
  [hep-th]} \BibitemShut {NoStop}%
\bibitem [{\citenamefont {Kinney}\ \emph {et~al.}(2007)\citenamefont {Kinney},
  \citenamefont {Maldacena}, \citenamefont {Minwalla},\ and\ \citenamefont
  {Raju}}]{Kinney:2005ej}%
  \BibitemOpen
  \bibfield  {author} {\bibinfo {author} {\bibfnamefont {J.}~\bibnamefont
  {Kinney}}, \bibinfo {author} {\bibfnamefont {J.~M.}\ \bibnamefont
  {Maldacena}}, \bibinfo {author} {\bibfnamefont {S.}~\bibnamefont {Minwalla}},
  \ and\ \bibinfo {author} {\bibfnamefont {S.}~\bibnamefont {Raju}},\ }\href
  {\doibase 10.1007/s00220-007-0258-7} {\bibfield  {journal} {\bibinfo
  {journal} {Commun. Math. Phys.}\ }\textbf {\bibinfo {volume} {275}},\
  \bibinfo {pages} {209} (\bibinfo {year} {2007})},\ \Eprint
  {http://arxiv.org/abs/hep-th/0510251} {arXiv:hep-th/0510251 [hep-th]}
  \BibitemShut {NoStop}%
\bibitem [{\citenamefont {Romelsberger}(2006)}]{Romelsberger:2005eg}%
  \BibitemOpen
  \bibfield  {author} {\bibinfo {author} {\bibfnamefont {C.}~\bibnamefont
  {Romelsberger}},\ }\href {\doibase 10.1016/j.nuclphysb.2006.03.037}
  {\bibfield  {journal} {\bibinfo  {journal} {Nucl. Phys.}\ }\textbf {\bibinfo
  {volume} {B747}},\ \bibinfo {pages} {329} (\bibinfo {year} {2006})},\ \Eprint
  {http://arxiv.org/abs/hep-th/0510060} {arXiv:hep-th/0510060 [hep-th]}
  \BibitemShut {NoStop}%
\bibitem [{\citenamefont {Hofman}\ and\ \citenamefont
  {Maldacena}(2008)}]{Hofman:2008ar}%
  \BibitemOpen
  \bibfield  {author} {\bibinfo {author} {\bibfnamefont {D.~M.}\ \bibnamefont
  {Hofman}}\ and\ \bibinfo {author} {\bibfnamefont {J.}~\bibnamefont
  {Maldacena}},\ }\href {\doibase 10.1088/1126-6708/2008/05/012} {\bibfield
  {journal} {\bibinfo  {journal} {JHEP}\ }\textbf {\bibinfo {volume} {05}},\
  \bibinfo {pages} {012} (\bibinfo {year} {2008})},\ \Eprint
  {http://arxiv.org/abs/0803.1467} {arXiv:0803.1467 [hep-th]} \BibitemShut
  {NoStop}%
\bibitem [{\citenamefont {Kutasov}\ \emph {et~al.}(2003)\citenamefont
  {Kutasov}, \citenamefont {Parnachev},\ and\ \citenamefont
  {Sahakyan}}]{Kutasov:2003iy}%
  \BibitemOpen
  \bibfield  {author} {\bibinfo {author} {\bibfnamefont {D.}~\bibnamefont
  {Kutasov}}, \bibinfo {author} {\bibfnamefont {A.}~\bibnamefont {Parnachev}},
  \ and\ \bibinfo {author} {\bibfnamefont {D.~A.}\ \bibnamefont {Sahakyan}},\
  }\href {\doibase 10.1088/1126-6708/2003/11/013} {\bibfield  {journal}
  {\bibinfo  {journal} {JHEP}\ }\textbf {\bibinfo {volume} {11}},\ \bibinfo
  {pages} {013} (\bibinfo {year} {2003})},\ \Eprint
  {http://arxiv.org/abs/hep-th/0308071} {arXiv:hep-th/0308071 [hep-th]}
  \BibitemShut {NoStop}%
\bibitem [{\citenamefont {Beem}\ and\ \citenamefont
  {Gadde}(2014)}]{Beem:2012yn}%
  \BibitemOpen
  \bibfield  {author} {\bibinfo {author} {\bibfnamefont {C.}~\bibnamefont
  {Beem}}\ and\ \bibinfo {author} {\bibfnamefont {A.}~\bibnamefont {Gadde}},\
  }\href {\doibase 10.1007/JHEP04(2014)036} {\bibfield  {journal} {\bibinfo
  {journal} {JHEP}\ }\textbf {\bibinfo {volume} {04}},\ \bibinfo {pages} {036}
  (\bibinfo {year} {2014})},\ \Eprint {http://arxiv.org/abs/1212.1467}
  {arXiv:1212.1467 [hep-th]} \BibitemShut {NoStop}%
\bibitem [{\citenamefont {Evtikhiev}(2018)}]{Evtikhiev:2017heo}%
  \BibitemOpen
  \bibfield  {author} {\bibinfo {author} {\bibfnamefont {M.}~\bibnamefont
  {Evtikhiev}},\ }\href {\doibase 10.1007/JHEP04(2018)120} {\bibfield
  {journal} {\bibinfo  {journal} {JHEP}\ }\textbf {\bibinfo {volume} {04}},\
  \bibinfo {pages} {120} (\bibinfo {year} {2018})},\ \Eprint
  {http://arxiv.org/abs/1708.08307} {arXiv:1708.08307 [hep-th]} \BibitemShut
  {NoStop}%
\bibitem [{\citenamefont {Intriligator}\ and\ \citenamefont
  {Seiberg}(1994)}]{Intriligator:1994sm}%
  \BibitemOpen
  \bibfield  {author} {\bibinfo {author} {\bibfnamefont {K.~A.}\ \bibnamefont
  {Intriligator}}\ and\ \bibinfo {author} {\bibfnamefont {N.}~\bibnamefont
  {Seiberg}},\ }\href {\doibase 10.1016/0550-3213(94)90215-1} {\bibfield
  {journal} {\bibinfo  {journal} {Nucl. Phys.}\ }\textbf {\bibinfo {volume}
  {B431}},\ \bibinfo {pages} {551} (\bibinfo {year} {1994})},\ \Eprint
  {http://arxiv.org/abs/hep-th/9408155} {arXiv:hep-th/9408155 [hep-th]}
  \BibitemShut {NoStop}%
\bibitem [{\citenamefont {Elitzur}\ \emph {et~al.}(1996)\citenamefont
  {Elitzur}, \citenamefont {Forge}, \citenamefont {Giveon},\ and\ \citenamefont
  {Rabinovici}}]{Elitzur:1995xp}%
  \BibitemOpen
  \bibfield  {author} {\bibinfo {author} {\bibfnamefont {S.}~\bibnamefont
  {Elitzur}}, \bibinfo {author} {\bibfnamefont {A.}~\bibnamefont {Forge}},
  \bibinfo {author} {\bibfnamefont {A.}~\bibnamefont {Giveon}}, \ and\ \bibinfo
  {author} {\bibfnamefont {E.}~\bibnamefont {Rabinovici}},\ }\href {\doibase
  10.1016/0550-3213(95)00564-1} {\bibfield  {journal} {\bibinfo  {journal}
  {Nucl. Phys.}\ }\textbf {\bibinfo {volume} {B459}},\ \bibinfo {pages} {160}
  (\bibinfo {year} {1996})},\ \Eprint {http://arxiv.org/abs/hep-th/9509130}
  {arXiv:hep-th/9509130 [hep-th]} \BibitemShut {NoStop}%
\bibitem [{\citenamefont {Bobev}\ and\ \citenamefont
  {Crichigno}(2017)}]{Bobev:2017uzs}%
  \BibitemOpen
  \bibfield  {author} {\bibinfo {author} {\bibfnamefont {N.}~\bibnamefont
  {Bobev}}\ and\ \bibinfo {author} {\bibfnamefont {P.~M.}\ \bibnamefont
  {Crichigno}},\ }\href {\doibase 10.1007/JHEP12(2017)065} {\bibfield
  {journal} {\bibinfo  {journal} {JHEP}\ }\textbf {\bibinfo {volume} {12}},\
  \bibinfo {pages} {065} (\bibinfo {year} {2017})},\ \Eprint
  {http://arxiv.org/abs/1708.05052} {arXiv:1708.05052 [hep-th]} \BibitemShut
  {NoStop}%
\bibitem [{Note2()}]{Note2}%
  \BibitemOpen
  \bibinfo {note} {There are two theories with 3 conserved currents with
  smaller $a$, but we do not find any evidence for the $SU(2)$
  symmetry.}\BibitemShut {Stop}%
\bibitem [{Note3()}]{Note3}%
  \BibitemOpen
  \bibinfo {note} {The theory with smaller $c$ than $H_0^*$ was also noticed by
  Sergio Benvenuti. We thank him for informing us on this.}\BibitemShut {Stop}%
\bibitem [{Note4()}]{Note4}%
  \BibitemOpen
  \bibinfo {note} {We might have anticipated this from the fact that $X$ was
  introduced to decouple ${\protect \rm Tr}\phi ^2$, but as we see in later
  examples $X$ might remain a generator.}\BibitemShut {Stop}%
\bibitem [{\citenamefont {Benvenuti}\ and\ \citenamefont
  {Giacomelli}(2017)}]{Benvenuti:2017lle}%
  \BibitemOpen
  \bibfield  {author} {\bibinfo {author} {\bibfnamefont {S.}~\bibnamefont
  {Benvenuti}}\ and\ \bibinfo {author} {\bibfnamefont {S.}~\bibnamefont
  {Giacomelli}},\ }\href {\doibase 10.1103/PhysRevLett.119.251601} {\bibfield
  {journal} {\bibinfo  {journal} {Phys. Rev. Lett.}\ }\textbf {\bibinfo
  {volume} {119}},\ \bibinfo {pages} {251601} (\bibinfo {year} {2017})},\
  \Eprint {http://arxiv.org/abs/1706.02225} {arXiv:1706.02225 [hep-th]}
  \BibitemShut {NoStop}%
\bibitem [{\citenamefont {Dolan}\ and\ \citenamefont
  {Osborn}(2003)}]{Dolan:2002zh}%
  \BibitemOpen
  \bibfield  {author} {\bibinfo {author} {\bibfnamefont {F.~A.}\ \bibnamefont
  {Dolan}}\ and\ \bibinfo {author} {\bibfnamefont {H.}~\bibnamefont {Osborn}},\
  }\href {\doibase 10.1016/S0003-4916(03)00074-5} {\bibfield  {journal}
  {\bibinfo  {journal} {Annals Phys.}\ }\textbf {\bibinfo {volume} {307}},\
  \bibinfo {pages} {41} (\bibinfo {year} {2003})},\ \Eprint
  {http://arxiv.org/abs/hep-th/0209056} {arXiv:hep-th/0209056 [hep-th]}
  \BibitemShut {NoStop}%
\bibitem [{\citenamefont {Gadde}\ \emph {et~al.}(2011)\citenamefont {Gadde},
  \citenamefont {Rastelli}, \citenamefont {Razamat},\ and\ \citenamefont
  {Yan}}]{Gadde:2010en}%
  \BibitemOpen
  \bibfield  {author} {\bibinfo {author} {\bibfnamefont {A.}~\bibnamefont
  {Gadde}}, \bibinfo {author} {\bibfnamefont {L.}~\bibnamefont {Rastelli}},
  \bibinfo {author} {\bibfnamefont {S.~S.}\ \bibnamefont {Razamat}}, \ and\
  \bibinfo {author} {\bibfnamefont {W.}~\bibnamefont {Yan}},\ }\href {\doibase
  10.1007/JHEP03(2011)041} {\bibfield  {journal} {\bibinfo  {journal} {JHEP}\
  }\textbf {\bibinfo {volume} {03}},\ \bibinfo {pages} {041} (\bibinfo {year}
  {2011})},\ \Eprint {http://arxiv.org/abs/1011.5278} {arXiv:1011.5278
  [hep-th]} \BibitemShut {NoStop}%
\bibitem [{\citenamefont {Gukov}(2016)}]{Gukov:2015qea}%
  \BibitemOpen
  \bibfield  {author} {\bibinfo {author} {\bibfnamefont {S.}~\bibnamefont
  {Gukov}},\ }\href {\doibase 10.1007/JHEP01(2016)020} {\bibfield  {journal}
  {\bibinfo  {journal} {JHEP}\ }\textbf {\bibinfo {volume} {01}},\ \bibinfo
  {pages} {020} (\bibinfo {year} {2016})},\ \Eprint
  {http://arxiv.org/abs/1503.01474} {arXiv:1503.01474 [hep-th]} \BibitemShut
  {NoStop}%
\bibitem [{\citenamefont {Gang}\ and\ \citenamefont
  {Yamazaki}(2018)}]{Gang:2018huc}%
  \BibitemOpen
  \bibfield  {author} {\bibinfo {author} {\bibfnamefont {D.}~\bibnamefont
  {Gang}}\ and\ \bibinfo {author} {\bibfnamefont {M.}~\bibnamefont
  {Yamazaki}},\ }\href {\doibase 10.1103/PhysRevD.98.121701} {\bibfield
  {journal} {\bibinfo  {journal} {Phys. Rev.}\ }\textbf {\bibinfo {volume}
  {D98}},\ \bibinfo {pages} {121701} (\bibinfo {year} {2018})},\ \Eprint
  {http://arxiv.org/abs/1806.07714} {arXiv:1806.07714 [hep-th]} \BibitemShut
  {NoStop}%
\end{thebibliography}%

\newpage

\begin{appendix}

\onecolumngrid

\section{APPENDIX: List of Candidate SCFTs}

Here we list the 70 candidate fixed points under investigation in the main letter. Tables \ref{tab:good1} and \ref{tab:bad1} list the central charges and R-charges of fields for the ``good" and ``bad" theories, respectively. The ``bad" theories are labeled as such when naively the $a$-maximization procedure holds, but there is a term in the index that indicates a sickness of the theory. We checked terms in the reduced index up to order $t^8$ to make this determination. We expect that there is something more to understand about these theories, and so have included their data here. 

There are generally multiple UV descriptions for the same fixed point. We've checked that these multiple descriptions are consistent with each other, and chosen a representative superpotential that flows to each fixed point. These superpotentials are listed in Tables \ref{tab:good2}  and \ref{tab:bad2}.  

Sometimes the fixed point theory has a flavor symmetry that is not manifest in any of the classical descriptions. This feature can be diagnosed with the $t^6$ term in the index, whose coefficient counts marginal operators minus conserved flavor currents. When the theory has a non-manifest $U(1)$, we cannot rule out the possibility that it mixes with the superconformal R-symmetry and changes the $a$-maximization answer. This is a common occurrence in the ``bad" theories. We take note of which of the ``good" theories have this feature in Table \ref{tab:good2}.

	
\def\arraystretch{1.64}

\begin{center}
	\begin{longtable}{|c|c||c|c|c|c|c|}
  \hline
    & $\left(a,c\right)$ &  $R(q)$ & $R(\widetilde{q})$ & $R(\phi)$ & $R(X_i)$ & $R(M_i)$ \\
    \hline \hline
    1 &  $\left(\frac{263}{768},\frac{271}{768}\right) \simeq (0.3424, 0.3529)$ & $\frac{11}{12}$ & $\frac{5}{12}$ & $\frac{1}{6}$ & $\frac{5}{3}$ & $1$   \\ \hline
   2 &  $\begin{array}{c} \left(   \frac{1465 \sqrt{1465}+81108}{397488},\frac{1051 \sqrt{1465}+29088}{198744}   \right) \\ \simeq (0.3451, 0.3488)\end{array}$ & $\frac{543-\sqrt{1465}}{546} $ & $\frac{75-\sqrt{1465}}{78} $ & $\frac{ \sqrt{1465}+3}{273}$ & $\frac{2 \left(270-\sqrt{1465}\right)}{273} $ & $ $   \\ \hline 
   3 & $\left(   \frac{711}{2048},\frac{807}{2048}   \right) \simeq (0.3472, 0.3940)$ & $\frac{3}{4}$ & $\frac{1}{4}$ & $\frac{1}{4}$ & $\frac{3}{2}$ & $\frac{5}{4},\frac{3}{4}$   \\ \hline
   4 & $\left(  \frac{43}{120},\frac{11}{30} \right) \simeq (0.3583, 0.3667)$ & $\frac{8}{15}$ & $\frac{14}{15}$ & $\frac{2}{15}$ & $\frac{26}{15}$ & $\frac{4}{5}$  \\ \hline
   5 & $\left(\frac{375}{1024},\frac{439}{1024}\right) \simeq (0.3662, 0.4287)$ & $\frac{3}{4}$ & $\frac{1}{4}$ & $\frac{1}{4}$ & $\frac{3}{2}$ & $\frac{5}{4}$,$\frac{3}{4}$,$\frac{3}{4}$   \\ \hline
   6 &  $\left(\frac{2211}{5488},\frac{1277}{2744}\right) \simeq (0.4029, 0.4654)$ & $\frac{4}{7}$ & $\frac{2}{7}$ & $\frac{2}{7}$ & $\frac{10}{7}$ & $\frac{8}{7},\frac{6}{7}$   \\ \hline
   7 &  $\left(\frac{14535}{35152},\frac{8535}{17576}\right) \simeq (0.4135, 0.4856)$ & $\frac{6}{13}$ & $\frac{4}{13}$ & $\frac{4}{13}$ & $\frac{18}{13}$ & $\frac{14}{13},\frac{12}{13},\frac{14}{13},\frac{12}{13}$   \\ \hline
   8 &  $\begin{array}{c} \left(\frac{7441 \sqrt{7441}+628560}{3072432},\frac{4606 \sqrt{7441}+348435}{1536216}\right) \\ \simeq (0.4135, 0.4854)\end{array}$ & $\frac{783 - 5 \sqrt{7441}}{759}$ & $\frac{147 + \sqrt{7441}}{759}$ & $\frac{147 + \sqrt{7441}}{759}$ & $\frac{2\left(612-\sqrt{7441} \right)}{759}$ & $\frac{359-\sqrt{7441}}{253},\frac{147+\sqrt{7441}}{253}$  \\ \hline
   9 &  $\left(\frac{285}{686},\frac{167}{343}\right) \simeq (0.4155, 0.4869)$ &
$ \frac{4}{7}$ & $\frac{2}{7}$ & $\frac{2}{7}$ & $\frac{10}{7}$ & $\frac{8}{7},\frac{6}{7},\frac{6}{7}$   \\ \hline
   10 &  $\left(\frac{924}{2197},\frac{1093}{2197}\right) \simeq (0.4206, 0.4975)$ & 
 $\frac{4}{13}$ & $\frac{6}{13}$ & $\frac{4}{13}$ & $ \frac{18}{13}$ & $\frac{10}{13},\frac{12}{13},\frac{14}{13},\frac{16}{13},\frac{12}{13}$   \\ \hline
   11 &  $\left(\frac{4 \left(896 \sqrt{7}+1665\right)}{38307},\frac{4036 \sqrt{7}+8355}{38307}\right) \simeq (0.4214, 0.4969)$ & $
 \frac{378-80 \sqrt{7}}{339} $ & $\frac{4\left(4 \sqrt{7}+15\right)}{339} $ & $\frac{4\left(4 \sqrt{7}+15\right)}{339} $ & $\frac{ -2 \left(16 \sqrt{7}-279\right)}{339}$ & $\begin{array}{c}\frac{-2\left(8 \sqrt{7}-83\right)}{113}  ,\frac{4\left(4 \sqrt{7}+15\right)}{113} ,\\ \frac{4\left(4 \sqrt{7}+15\right)}{113}\end{array}$   \\ \hline
   12 &  $\left(\frac{7587}{17576},\frac{2277}{4394}\right) \simeq (0.4317, 0.5182)$ & $
 \frac{6}{13}$ & $ \frac{4}{13}$ & $  \frac{4}{13} $& $\frac{18}{13}$ & $
 \frac{14}{13},\frac{12}{13},\frac{14}{13},\frac{10}{13},\frac{12}{13}$   \\ \hline
   13 &  $\left(\frac{339}{784},\frac{97}{196}\right) \simeq (0.4324, 0.4949)$ & $
 \frac{1}{2}$ & $\frac{5}{14}$ & $\frac{2}{7}$ & $\frac{10}{7}$ & $1,\frac{6}{7},\frac{8}{7}$  \\ \hline
   14 &  $\begin{array}{c} \left(\frac{5665 \sqrt{5665}+162189}{1359456},\frac{5903 \sqrt{5665}+262863}{1359456}\right) \\\simeq (0.4329, 0.5202)\end{array} $ & $
 \frac{ 5 \sqrt{5665}-27}{714}$ & $\frac{291-\sqrt{5665}}{714} $ & $\frac{291-\sqrt{5665}}{714} $ & $\frac{\sqrt{5665}+423}{357}$ & $\begin{array}{cc} \frac{\sqrt{5665}+185}{238} ,\frac{291-\sqrt{5665}}{238},\\
    \frac{397-3 \sqrt{5665}}{238}\end{array}$   \\ \hline
  15 &   $\left(\frac{15423}{35152},\frac{9317}{17576}\right) \simeq (0.4388, 0.5301)$ & $
 \frac{4}{13}$ & $\frac{6}{13}$ & $\frac{4}{13}$ & $\frac{18}{13}$& $\frac{10}{13},\frac{12}{13},\frac{14}{13},\frac{16}{13},\frac{12}{13},\frac{10}{13}$  \\ \hline
   16 &  $\begin{array}{c} \left(\frac{24817 \sqrt{24817}+1456776}{12144432},\frac{13666 \sqrt{24817}+1101111}{6072216}\right)\\ \simeq (0.4419, 0.5359)\end{array} $ & $
 \frac{5 \sqrt{24817}-27}{1509}$ & $  \frac{609-\sqrt{24817}}{1509}$ & $\frac{609-\sqrt{24817}}{1509}$ & $\frac{2 \left(\sqrt{24817}+900\right)}{1509}$ & $\begin{array}{c}\frac{ \sqrt{24817}+397}{503},\frac{ 609-\sqrt{24817}}{503},\\
 \frac{609-\sqrt{24817}}{503} ,\frac{821-3 \sqrt{24817}}{503} \end{array}$  \\ \hline
   17 &  $\left(\frac{1221}{2744},\frac{1417}{2744}\right) \simeq (0.4450, 0.5164)$ &
 $\frac{1}{2}$ & $\frac{5}{14}$ & $\frac{2}{7}$ & $\frac{10}{7}$ & $1,\frac{6}{7},\frac{8}{7},\frac{6}{7}$   \\ \hline
   18 &  $\left(\frac{97 \sqrt{97}+423}{3072},\frac{113 \sqrt{97}+471}{3072}\right) \simeq (0.4487, 0.5156)$ & $
 \frac{ 123-7 \sqrt{97}}{96}$ & $\frac{ 45-\sqrt{97}}{96}$ & $\frac{\sqrt{97}+3}{48} $ & $\frac{45-\sqrt{97}}{24} $ & $1,\frac{\sqrt{97}+3}{16} $  \\ \hline
   19 &  $\left(\frac{19 \sqrt{19}- 72}{24},\frac{5(4 \sqrt{19}-15)}{24}\right) \simeq (0.4508, 0.5074)$ & 
 $\frac{ 7-\sqrt{19}}{4}$ & $\frac{27-5 \sqrt{19}}{12}$ & $\frac{\sqrt{19}-3}{6} $ & $\frac{9-\sqrt{19}}{3}$ & $\begin{array}{c}\frac{2\left(\sqrt{19}-3\right)}{3} ,\frac{2\left(6- \sqrt{19}\right)}{3},\\\frac{\sqrt{19}-3}{2}\end{array}$  \\ \hline
   20 &  $\left(\frac{621}{1372},\frac{2925}{5488}\right) \simeq (0.4526, 0.5330)$ & $
 \frac{1}{2} $ & $ \frac{5}{14} $ & $ \frac{2}{7} $ & $ \frac{10}{7} $ & $1,\frac{6}{7},\frac{8}{7},\frac{5}{7}$   \\ \hline
   21 &  $\left(\frac{927}{2048},\frac{1023}{2048}\right) \simeq (0.4526, 0.4995)$ & $
 \frac{1}{2} $ & $ \frac{1}{2} $ & $ \frac{1}{4} $ & $\frac{3}{2}$& $1$   \\ \hline
   22 &  $\left(\frac{601 \sqrt{601}+15012}{65712},\frac{430 \sqrt{601}+5841}{32856}\right) \simeq (0.4527, 0.4986)$ & $
 \frac{105-2 \sqrt{601}}{111} $& $\frac{105-2 \sqrt{601}}{111}$ & $\frac{\sqrt{601}+3 }{111} $ & $\frac{-2 \left(\sqrt{601}-108\right)}{111} $ &   \\ \hline
   23 &  $\left(\frac{11}{24},\frac{1}{2}\right) \simeq (0.4583, 0.5000)$ & $
 \frac{5}{9}$ & $ \frac{5}{9} $ & $ \frac{2}{9} $ & $\frac{14}{9}$ & $ \frac{8}{9}$   \\ \hline
   24 &  $\left(\frac{2553}{5488},\frac{3043}{5488}\right) \simeq (0.4652, 0.5545)$ & $
 \frac{1}{2} $ & $ \frac{5}{14}  $ & $ \frac{2}{7} $ & $ \frac{10}{7}$ & $ 1,\frac{6}{7},\frac{8}{7},\frac{5}{7},\frac{6}{7} $  \\ \hline
   25 &  $\left(\frac{483}{1024},\frac{547}{1024}\right) \simeq (0.4717, 0.5342)$ & $
 \frac{1}{2} $ & $ \frac{1}{2} $ & $ \frac{1}{4} $ & $ \frac{3}{2}$ & $1,\frac{3}{4}$  \\ \hline
   26 &  $\left(\frac{352 \sqrt{22}+1251}{6144},\frac{416 \sqrt{22}+1347}{6144}\right) \simeq (0.4723, 0.5368)$ & $
 \frac{2 \sqrt{22}+3}{24} $ & $\frac{21-2 \sqrt{22} }{24} $& $\frac{1}{4}$ & $\frac{3}{2}$& $1,\frac{9-\sqrt{22}}{6} $  \\ \hline
   27 &  $\left(\frac{61 \sqrt{61}-441}{75} ,\frac{ 127 \sqrt{61}-912}{150}\right) \simeq (0.4723, 0.5327)$ & $
 \frac{39-4 \sqrt{61}}{15}$  & $\frac{39-4 \sqrt{61}}{15} $ & $\frac{2\left(\sqrt{61}-6\right)}{15}$  & $ \frac{2\left(27-2 \sqrt{61}\right)}{15}$ & $\frac{2\left(\sqrt{61}-6\right)}{5} $  \\
        28 &  $\left( 0.4727,0.5351 \right)$ &  $
 0.5258 $ & $ 0.5009 $  & $ 0.2433$  & $  1.513 $ & $ 0.7051 $   \\ \hline
   29 &  $\left(\frac{1005}{2048},\frac{1165}{2048}\right) \simeq (0.4907, 0.5688)$ & $\frac{1}{2} $ & $\frac{1}{2}$ & $\frac{1}{4}$ & $\frac{3}{2}$ & $1,\frac{3}{4},\frac{3}{4}$  \\
\hline
   30 &  $\left(\frac{44 \sqrt{22}+171}{768} ,\frac{13  \left(4 \sqrt{22}+15\right)}{768}\right) \simeq (0.4914, 0.5715)$ & 
 $\frac{2 \sqrt{22}+3}{24} $ & $\frac{21-2 \sqrt{22}}{24} $ & $\frac{1}{4}$ & $\frac{3}{2}$ & $1,\frac{3}{4},\frac{9-\sqrt{22}}{6} $  \\ \hline
   31 &  $\left(\frac{89 \sqrt{\frac{89}{17}}-180}{48},\frac{44 \sqrt{\frac{89}{17}}-87}{24}\right) \simeq (0.4925, 0.5698)$ & 
$ \frac{2 \sqrt{\frac{89}{17}}-3}{3}$ & $ \frac{2 \sqrt{\frac{89}{17}}-3}{3}$ & $\frac{3-\sqrt{\frac{89}{17}}}{3} $ & $\frac{2 \sqrt{\frac{89}{17}}}{3}$ & $3-\sqrt{\frac{89}{17}},3-\sqrt{\frac{89}{17}}$  \\ \hline
   32 &  $\left( 0.4927,0.5714  \right)$ & $ 0.5129 $& $ 0.5326$ & $ 0.2386$ & $1.523$ & $0.7159,0.6962$   \\ \hline 
   33 &  $\left(\frac{261}{512},\frac{309}{512}\right) \simeq (0.5098, 0.6035)$ &
 $\frac{1}{2}$ & $\frac{1}{2}$ & $\frac{1}{4} $ & $\frac{3}{2}$ & $1,\frac{3}{4},\frac{3}{4},\frac{3}{4}$ \\ \hline
   34 &  $\begin{array}{c} \left(\frac{553 \sqrt{553}-7047}{11616},\frac{575 \sqrt{553}-6453}{11616}\right) \\
   \simeq (0.5129, 0.6085)\end{array}$ & $ \frac{\sqrt{553}-6}{33} $  & $ \frac{\sqrt{553}-6 }{33} $ & $ \frac{39-\sqrt{553}}{66}$  & $\frac{\sqrt{553}+27}{33} $ & $ \frac{39-\sqrt{553}}{22} ,\frac{39-\sqrt{553}}{22},\frac{39-\sqrt{553}}{22} $  \\
   \hline
  \caption{Candidate SCFTs that pass all tests they've been subjected to, ordered by increasing $a$ central charge.} 
    	\label{tab:good1}
\end{longtable}
\end{center}

\begin{center}
	\begin{longtable}{|c||c|c|c|}
  \hline
   & Superpotential $W$ & Notes  &  $\begin{array}{c} \text{Non-manifest} \\ \text{symmetry?}  \end{array}$ \\  \hline  \hline
    1 & $X_1\tr \phi^2 + M_1 \phi  \tilde{q}^2+M_1^2+q^2 \phi $   &  $H_0^*$  &  \\ \hline
    2 & $X_1\tr \phi^2 +  q^2 \phi  $  & $\CT_0$ & \\ \hline
    3 & $X_1\tr \phi^2 +  M_1 \phi  \tilde{q}^2+M_2 q \phi  \tilde{q}+q^2 \tilde{q}^2+M_1 M_2  $ & $\CT_\mu$ & \\ \hline
    4 & $X_1\tr \phi^2 +  \phi  \tilde{q}^2+M_1 q^2 \phi  $ & $H_0$  &  \\ \hline
    5 & $X_1\tr \phi^2 +  M_1 \phi  \tilde{q}^2+M_2 q \phi  \tilde{q}+q^2 \tilde{q}^2+M_1 M_2+M_1 M_3  $ &  & yes \\ \hline
    6 & $X_1\tr \phi^2 +  M_1 \phi  \tilde{q}^2+M_2 q \phi  \tilde{q}+M_1 q \tilde{q}+M_1 M_2  $ &  & yes \\ \hline
    7 & $X_1\tr \phi^2 +   M_2 q \phi  \tilde{q}+M_1 \phi  \tilde{q}^2+q^3 \phi  \tilde{q}+M_1 M_2+M_2 M_3+M_1 M_4 $ & & \\ \hline
    8 & $X_1\tr \phi^2 +  M_1 \phi  \tilde{q}^2+M_2 q \phi  \tilde{q}+M_1 M_2  $ & & \\ \hline
    9 & $X_1\tr \phi^2 +  M_1 \phi  \tilde{q}^2+M_2 q \phi  \tilde{q}+M_1 q \tilde{q}+M_1 M_2+M_1 M_3 $  & & yes\\ \hline
    10 & $X_1\tr \phi^2 + M_2 q \phi  \tilde{q}+M_1 \phi  \tilde{q}^2+M_4 q \tilde{q}+M_3 q^2 \phi +M_2 M_3+M_1 M_4+M_3 M_5 $ & & yes \\ \hline
    11 & $X_1\tr \phi^2 +  M_1 \phi  \tilde{q}^2+M_2 q \phi  \tilde{q}+M_1 M_2+M_1 M_3 $ & & yes \\ \hline
    12 & $X_1\tr \phi^2 +  M_2 q \phi  \tilde{q}+M_1 \phi  \tilde{q}^2+q^3 \phi  \tilde{q}+M_4 q^2 \phi +M_1 M_2+M_2 M_3+M_1 M_5 $ & & yes \\ \hline
    13 & $X_1\tr \phi^2 + M_1 \phi  \tilde{q}^2+M_2 q \phi  \tilde{q}+M_3 q \tilde{q}+M_1^2+M_2 M_3  $ & & \\ \hline
    14 & $X_1\tr \phi^2 +  M_2 q \phi  \tilde{q}+M_1 \phi  \tilde{q}^2+M_3 q^2 \phi +M_1 M_2 $  & & \\ \hline
    15 & $X_1\tr \phi^2 +  M_2 q \phi  \tilde{q}+M_1 \phi  \tilde{q}^2+M_4 q \tilde{q}+M_3 q^2 \phi +M_2 M_3+M_1 M_4+M_3 M_5+M_4 M_6 $  & & yes \\ \hline
    16 & $X_1\tr \phi^2 +  M_2 q \phi  \tilde{q}+M_1 \phi  \tilde{q}^2+M_4 q^2 \phi +M_1 M_2+M_1 M_3 $ & & yes \\ \hline
    17 & $X_1\tr \phi^2 +  M_1 \phi  \tilde{q}^2+M_2 q \phi  \tilde{q}+M_3 q \tilde{q}+M_1^2+M_2 M_3+M_3 M_4 $ & & yes \\  \hline
    18 & $X_1\tr \phi^2 + M_1 \phi  \tilde{q}^2+M_2 q \phi  \tilde{q}+M_1^2  $ & & \\ \hline
    19 & $X_1\tr \phi^2 + M_1 \phi  \tilde{q}^2+M_3 q \phi  \tilde{q}+M_1 q \tilde{q}+M_1 M_2  $ & & yes\\ \hline
    20 & $X_1\tr \phi^2 + M_2 q \phi  \tilde{q}+M_1 \phi  \tilde{q}^2+M_3 q \tilde{q}+M_4 q^2 \phi +M_1^2+M_2 M_3   $ & & \\ \hline
    21 & $X_1\tr \phi^2 +  M_1 q \tilde{q}+M_1^2 $ & $H_1^*$ &  \\ \hline
    22 & $X_1\tr \phi^2 $ & $\widehat{\CT}$ & \\ \hline
    23 & $X_1\tr \phi^2 +  M_1 q \tilde{q}  $ & $H_1$ & \\ \hline
    24 & $X_1\tr \phi^2 +  M_2 q \phi  \tilde{q}+M_1 \phi  \tilde{q}^2+M_3 q \tilde{q}+M_4 q^2 \phi +M_1^2+M_2 M_3+M_3 M_5  $ & &   yes \\ \hline
    25 & $X_1\tr \phi^2 + M_2 q \phi  \tilde{q}+M_1 q \tilde{q}+M_1^2   $ & & \\  \hline
    26 & $X_1\tr \phi^2 +  M_1 q \tilde{q}+M_2 q^2 \phi +M_1^2  $ & & \\ \hline
    27 & $X_1\tr \phi^2 + M_1 q \phi  \tilde{q}   $ & & \\ \hline
    28 & $X_1\tr \phi^2 +  M_1 q^2 \phi $ & & \\ \hline
    29 & $X_1\tr \phi^2 + M_2 \phi  \tilde{q}^2+M_1 q \tilde{q}+M_3 q^2 \phi +M_1^2  $  & & \\ \hline
    30 & $X_1\tr \phi^2 +   M_2 q \phi  \tilde{q}+M_1 q \tilde{q}+M_3 q^2 \phi +M_1^2 $ & & \\ \hline
    31 & $X_1\tr \phi^2 +   M_1 \phi  \tilde{q}^2+M_2 q^2 \phi $  & & \\ \hline
    32 & $X_1\tr \phi^2 +   M_2 \phi  \tilde{q}^2+M_1 q \phi  \tilde{q} $ & & \\ \hline
    33 & $X_1\tr \phi^2 +  M_4 q \phi  \tilde{q}+M_2 \phi  \tilde{q}^2+M_1 q \tilde{q}+M_3 q^2 \phi +M_1^2  $ & & \\ \hline
    34 & $X_1\tr \phi^2 +  M_3 q \phi  \tilde{q}+M_1 \phi  \tilde{q}^2+M_2 q^2 \phi $ & & \\
    \hline 
  \caption{Corresponding superpotentials to Table \ref{tab:good1}.
  The last column indicates the existence of an accidental symmetry in the IR which is not manifest in the UV.} 
    	\label{tab:good2}
\end{longtable}
\end{center}

\begin{center}
	\begin{longtable}{|c|c||c|c|c|c|c|}
  \hline
    & $\left(a,c\right)$ &  $R(q)$ & $R(\widetilde{q})$ & $R(\phi)$ & $R(X_i)$ & $R(M_i)$  \\
    \hline \hline
 1 & $ \left(\frac{141}{1000},\frac{83}{500}\right) \simeq (0.1410, 0.1660) $ & $ \frac{2}{5} $ & 0 & $ \frac{2}{5} $ & $\frac{6}{5},\frac{8}{5}$ & $\frac{8}{5},\frac{6}{5},\frac{4}{5},\frac{6}{5}$   \\ \hline
 2 & $ \left(\frac{63}{400},\frac{39}{200}\right)\simeq (0.1575, 0.1950)$  & $ \frac{2}{5}$ & 0 & $ \frac{2}{5} $ & $\frac{6}{5},\frac{8}{5}$ & $\frac{8}{5},\frac{6}{5},\frac{4}{5}$    \\ \hline
 3 & $ \left(\frac{87}{500},\frac{28}{125}\right)\simeq (0.1740, 0.2240)$  & $\frac{2}{5}$ & 0 & $\frac{2}{5} $& $\frac{6}{5},\frac{8}{5}$& $\frac{8}{5},\frac{6}{5},\frac{4}{5},\frac{4}{5}$    \\ \hline
 4 & $\left(\frac{381}{2000},\frac{253}{1000}\right)\simeq (0.1905, 0.2530)$ & $\frac{2}{5}$ & 0 & $\frac{2}{5} $ & $\frac{6}{5},\frac{8}{5}$ & $\frac{8}{5},\frac{6}{5},\frac{4}{5},\frac{4}{5},\frac{4}{5}$    \\ \hline
 5 & $\left(\frac{417}{2048},\frac{449}{2048}\right)\simeq (0.2036, 0.2192)$ & $\frac{7}{8}$ & $ \frac{1}{8} $ & $ \frac{1}{4} $ & $\frac{3}{2}$ & $1,\frac{3}{2}$    \\ \hline
 6 & $\left(\frac{57}{256},\frac{65}{256}\right)\simeq (0.2227, 0.2539)$  & $\frac{7}{8}$ & $ \frac{1}{8} $ & $\frac{1}{4} $ & $\frac{3}{2}$ & $ 1,\frac{3}{2},\frac{3}{4}$    \\ \hline
 7 & $\left(\frac{169}{648},\frac{89}{324}\right)\simeq (0.2608, 0.2747)$ & $\frac{8}{9} $ & $ \frac{2}{9} $ & $ \frac{2}{9} $ & $\frac{14}{9}$ & $\frac{4}{3},\frac{2}{3},\frac{4}{3}$  \\ \hline
 8 & $\left(\frac{13}{48},\frac{7}{24}\right)\simeq(0.2708, 0.2917)$  & $\frac{8}{9}$ & $\frac{2}{9}$ & $\frac{2}{9}$ & $\frac{14}{9}$ &  $\frac{4}{3},\frac{2}{3},\frac{4}{3},\frac{8}{9}$   \\ \hline
 9 & $\left(\frac{7}{24},\frac{1}{3}\right)\simeq(0.2917, 0.3333)$ & $\frac{1}{2}$ & $\frac{1}{6}$ & $\frac{1}{3}$ & $ \frac{4}{3},\frac{4}{3},\frac{4}{3}$ & $\frac{4}{3},\frac{2}{3},1$   \\ \hline
 10 & $\left(\frac{21}{64},\frac{23}{64}\right)\simeq (0.3281, 0.3594)$ & $ \frac{3}{4} $ & $ \frac{1}{4} $ & $\frac{1}{4}$ & $\frac{3}{2}$ & $\frac{5}{4},\frac{3}{4},\frac{5}{4}$    \\ \hline
 11 & $\left(\frac{3}{8},\frac{7}{16}\right)\simeq (0.3750, 0.4375)$ &  $ \frac{1}{3} $ & $\frac{1}{3}$ & $\frac{1}{3}$ & $\frac{4}{3},\frac{4}{3}$ & $1$    \\ \hline
 12 & $\left(\frac{2073}{5488},\frac{1159}{2744}\right)\simeq (0.3777, 0.4224)$ & $\frac{4}{7}$  & $\frac{2}{7}$ & $\frac{2}{7}$ & $\frac{10}{7} $ & $\frac{8}{7},\frac{6}{7},\frac{8}{7},\frac{8}{7} $    \\ \hline
 13 & $\left(\frac{5 \sqrt{5}+7}{48},\frac{\sqrt{5}}{8}+\frac{1}{6}\right)\simeq (0.3788, 0.4462)$ & $\frac{4-\sqrt{5}}{6} $ & $ \frac{\sqrt{5}}{6}$ & $\frac{1}{3}$ & $\frac{4}{3},\frac{4}{3}$ & $\frac{5-\sqrt{5}}{3} ,1$   \\ \hline
 14 & $\left(\frac{13647}{35152},\frac{7753}{17576}\right)\simeq (0.3882, 0.4411)$  & $\frac{6}{13}$ & $\frac{4}{13} $& $ \frac{4}{13}$ & $\frac{18}{13}$& $ \frac{14}{13},\frac{12}{13},\frac{14}{13},\frac{16}{13}$    \\ \hline
 15 & $\left(\frac{17 \sqrt{17}+79}{384} ,\frac{21 \sqrt{17}+83}{384}\right)\simeq (0.3883, 0.4416)$ & $\frac{37-5 \sqrt{17}}{36} $ & $ \frac{\sqrt{17}+7}{36}  $ & $ \frac{\sqrt{17}+7}{36} $ & $\frac{29-\sqrt{17}}{18} $ & $\begin{array}{c}  \frac{17-\sqrt{17}}{12} ,\frac{\sqrt{17}+7}{12} ,\\\frac{\sqrt{17}+7}{9} ,\frac{17-\sqrt{17}}{12} \end{array}$  \\ \hline
 16 & $\left(\frac{153}{392},\frac{87}{196}\right)\simeq (0.3903, 0.4439)$  & $\frac{4}{7}$ & $ \frac{2}{7}$ & $\frac{2}{7}$ & $\frac{10}{7}$ & $\frac{8}{7},\frac{6}{7},\frac{8}{7}$    \\ \hline
 17 & $\left(\frac{1737}{4394},\frac{3981}{8788}\right)\simeq (0.3953, 0.4530)$ & $\frac{6}{13}$ & $\frac{4}{13}$ & $\frac{4}{13}$ & $\frac{18}{13}$ & $\frac{14}{13},\frac{12}{13},\frac{14}{13},\frac{16}{13},\frac{12}{13}$    \\ \hline
 18 & $\left(\frac{65 \sqrt{65}+407}{2352},\frac{18 \sqrt{65}+121}{588} \right)\simeq (0.3959, 0.4526)$  & $ \frac{71-5 \sqrt{65}}{63} $ & $ \frac{\sqrt{65}+11}{63} $ & $ \frac{\sqrt{65}+11}{63}$ & $\frac{-2 \left(\sqrt{65}-52\right)}{63} $ & $ \frac{31-\sqrt{65}}{21} ,\frac{\sqrt{65}+11}{21} ,\frac{4\left(\sqrt{65}+11\right)}{63} $    \\ \hline
 19 & $ \left(\frac{14145}{35152},\frac{8171}{17576}\right)\simeq (0.4024, 0.4649)$ &$  \frac{4}{13} $  & $\frac{6}{13}$ & $\frac{4}{13}$ & $\frac{18}{13}$ & $\frac{10}{13},\frac{12}{13},\frac{14}{13},\frac{16}{13},\frac{16}{13},\frac{12}{13}$    \\ \hline
 20 & $\left(\frac{175}{432},\frac{101}{216}\right)\simeq (0.4051, 0.4676)$ & $ \frac{14}{27}$ & $ \frac{8}{27} $ & $ \frac{8}{27}$ &$\frac{38}{27}$ & $\frac{10}{9},\frac{8}{9},\frac{32}{27},\frac{8}{9}$   \\ \hline
 21 & $\left(\frac{7143}{17576},\frac{4163}{8788}\right) \simeq (0.4064, 0.4737)$ & $\frac{6}{13}$ & $ \frac{4}{13}$ & $\frac{4}{13}$ & $\frac{18}{13}$ & $\frac{14}{13},\frac{12}{13},\frac{14}{13}$    \\ \hline
 22 & $\begin{array}{c} \left(\frac{1969 \sqrt{1969}+104103}{470400},\frac{2669 \sqrt{1969}+106203}{470400}\right)\\ \simeq (0.4070, 0.4775)\end{array}$  & $ \frac{81-\sqrt{1969}}{84} $ & $  \frac{\sqrt{1969}+87}{420}$ & $\frac{\sqrt{1969}+87}{420} $ & $ \frac{333-\sqrt{1969}}{210}$ & $\begin{array}{c} \frac{193-\sqrt{1969}}{140},\frac{\sqrt{1969}+87}{140} ,\\\frac{193-\sqrt{1969}}{140} \end{array}$     \\ \hline
 23 & $\begin{array}{c}  \left(\frac{233 \sqrt{233}+3503}{17328},\frac{7 \left(9 \sqrt{233}+157\right)}{4332}\right)\\ \simeq (0.4074, 0.4757)\end{array}$ & $ \frac{5 \sqrt{233}+7}{171} $ & $ \frac{67-\sqrt{233}}{171}$ &  $\frac{67-\sqrt{233}}{171} $ & $\frac{2 \left(\sqrt{233}+104\right)}{171} $ & $ \begin{array}{c} \frac{\sqrt{233}+47}{57} ,\frac{67-\sqrt{233}}{57} ,\\ \frac{\sqrt{233}+47}{57} , \frac{ -4 \left(\sqrt{233}-67\right)}{171}, \\ \frac{29-\sqrt{233}}{19}\end{array} $    \\ \hline
 24 & $\left(\frac{13 \sqrt{65}+95}{480},\frac{3 \sqrt{65}+23}{96} \right)\simeq (0.4163, 0.4915)$ & $ \frac{\sqrt{65}+1}{18}  $ & $ \frac{35-\sqrt{65}}{90}$   & $ \frac{35-\sqrt{65}}{90}$ & $\frac{\sqrt{65}+55}{45}$ & $\begin{array}{c} \frac{\sqrt{65}+25}{30} ,\frac{35-\sqrt{65}}{30} ,\\\frac{-2\left(\sqrt{65}-35\right)}{45}  ,\frac{15-\sqrt{65}}{10}\end{array}$     \\ \hline
 25 & $\left(\frac{144}{343},\frac{1299}{2744}\right)\simeq (0.4198, 0.4734)$ & $\frac{1}{2} $ & $ \frac{5}{14}$ & $ \frac{2}{7}$ & $\frac{10}{7}$ & $1,\frac{6}{7},\frac{8}{7},\frac{8}{7}$    \\ \hline
 26 & $ \left(\frac{14925}{35152},\frac{8899}{17576}\right)\simeq (0.4246, 0.5063)$ & $\frac{6}{13}$ & $\frac{4}{13}$ & $\frac{4}{13}$ & $\frac{18}{13} $ & $ \frac{14}{13},\frac{12}{13},\frac{14}{13},\frac{10}{13}$     \\ \hline
 27 & $\begin{array}{c} \left(\frac{20665 \sqrt{20665}+1140192}{9676848},\frac{10108 \sqrt{20665}+997095}{4838424}\right)\\ 
 \simeq (0.4248, 0.5064)\end{array}$ & $\frac{5 \sqrt{20665}-81}{1347} $ & $\frac{555-\sqrt{20665}}{1347}$ & $\frac{555-\sqrt{20665}}{1347} $ & $ \frac{2 \left(\sqrt{20665}+792\right)}{1347}$ & $ \begin{array}{c} \frac{\sqrt{20665}+343}{449} ,\frac{555-\sqrt{20665}}{449} , \\ \frac{\sqrt{20665}+343}{449},\frac{767-3 \sqrt{20665}}{449} \end{array}$    \\ \hline
 28 & $ \left(\frac{873}{2048},\frac{969}{2048}\right)\simeq (0.4263, 0.4731)$ & $ \frac{5}{8} $ & $ \frac{3}{8} $ & $ \frac{1}{4} $ &  $\frac{3}{2} $ & 1     \\ \hline
 29 & $ \left(\frac{11 \sqrt{11}-16}{48},\frac{12\sqrt{11}-17}{48}\right)\simeq (0.4267, 0.4750)$ & $\frac{34-7 \sqrt{11}}{18} $ & $\frac{10-\sqrt{11}}{18}$ & $\frac{\sqrt{11}-1}{9} $ & $\frac{-2 \left(\sqrt{11}-10\right)}{9} $ & $ 1,\frac{4\left(\sqrt{11}-1\right)}{9} $    \\ \hline
 30 & $ \begin{array}{c}  \left(\frac{697 \sqrt{697}+19764}{88752},\frac{499 \sqrt{697}+7560}{44376}\right)\\ \simeq (0.4300, 0.4672)\end{array}$ & $\frac{83-\sqrt{697}}{86} $ &  $ \frac{243-5 \sqrt{697}}{258} $ & $ \frac{\sqrt{697}+3}{129} $& $ \frac{-2 \left(\sqrt{697}-126\right)}{129} $ & $\frac{4\left(\sqrt{697}+3\right)}{129} ,\frac{2\left(123-2 \sqrt{697}\right)}{129}$    \\ \hline
 31 & $ \begin{array}{c}  \left(\frac{103 \sqrt{1339}+9477}{30576},\frac{152 \sqrt{1339}+9477}{30576}\right) \\\simeq (0.4332, 0.4919)\end{array} $ & $\frac{6-\sqrt{\frac{103}{13}}}{6}$ & $\frac{18-\sqrt{\frac{103}{13}}}{42}$ & $ \frac{\sqrt{1339}+39}{273}$  & $\frac{-2 \left(\sqrt{1339}-234\right)}{273} $ & 1   \\ \hline
 32 & $\left(\frac{345}{784},\frac{401}{784}\right) \simeq(0.4401, 0.5115)$ & $\frac{1}{2}$  & $\frac{5}{14}$ & $\frac{2}{7}$ & $\frac{10}{7} $ & $1,\frac{6}{7},\frac{8}{7},\frac{5}{7},\frac{8}{7}$    \\ \hline
 33 & $ \left(\frac{113 \sqrt{113}-855}{784} ,\frac{53 \sqrt{113}-375}{392}\right)\simeq (0.4416, 0.4806)$ & $\frac{\sqrt{113}-1}{14} $ & $ \frac{5 \sqrt{113}-33}{42}$  & $\frac{15-\sqrt{113}}{21}  $ & $\frac{2 \left(\sqrt{113}+6\right)}{21} $ & $\frac{-4 \left(\sqrt{113}-15\right)}{21} $   \\ \hline
 34 &  $\left(\frac{57}{128},\frac{65}{128}\right) \simeq (0.4453, 0.5078)$ &
 $\frac{5}{8}$ & $\frac{3}{8}$ & $\frac{1}{4}$ & $\frac{3}{2}$& $1,1,\frac{3}{4}$ \\ \hline
 35 & $ \left(\frac{6349}{13872},\frac{3523}{6936}\right)\simeq (0.4577, 0.5079)$  & $ \frac{12}{17}$ & $\frac{26}{51}$ & $\frac{10}{51}$ & $\frac{82}{51}$ & $\frac{40}{51},\frac{40}{51}$   \\ \hline
 36 & $ \left(\frac{125}{48},\frac{59}{24}\right) \simeq (2.604, 2.458)$  & $\frac{2}{3} $ & $-\frac{4}{3} $ & $ \frac{2}{3}$  & $\frac{2}{3},2,\frac{4}{3} $ &  $ 4,2,0,\frac{8}{3} $    \\ 
   \hline
  \caption{Candidate SCFTs whose indices don't pass all tests, ordered by increasing $a$ central charge.} 
    	\label{tab:bad1}
  \end{longtable}
\end{center}


\begin{center}
    	\begin{longtable}{|c||c|c|}
  \hline
   & Superpotential $W$ & Notes  \\  \hline  \hline
 1 & $X_1 \phi ^2+ M_2 q \phi  \tilde{q}+M_1 \phi  \tilde{q}^2+M_1 q \tilde{q}+X_2 \phi  \tilde{q}^2+M_3 q^2 \phi +M_2 M_3+M_3 M_4$ &   \\ \hline
 2 & $X_1 \phi ^2+M_2 q \phi  \tilde{q}+M_1 \phi  \tilde{q}^2+M_1 q \tilde{q}+X_2 \phi  \tilde{q}^2+M_3 q^2 \phi +M_2 M_3$ &  \\ \hline
 3 &$ X_1 \phi ^2+M_2 q \phi  \tilde{q}+M_1 \phi  \tilde{q}^2+M_1 q \tilde{q}+X_2 \phi  \tilde{q}^2+M_3 q^2 \phi +M_2 M_3+M_2 M_4$ &   \\ \hline
 4 & $X_1 \phi ^2 +M_2 q \phi  \tilde{q}+M_1 \phi  \tilde{q}^2+M_1 q \tilde{q}+X_2 \phi  \tilde{q}^2+M_3 q^2 \phi +M_4 X_1+M_2 M_3+M_2 M_5$ &  \\ \hline
 5 & $X_1 \phi ^2+M_2 \phi  \tilde{q}^2+M_1 q \tilde{q}+M_1^2+q^2 \phi $ &  $T_M$   \\ \hline
 6 & $X_1 \phi ^2+M_3 q \phi  \tilde{q}+M_2 \phi  \tilde{q}^2+M_1 q \tilde{q}+M_1^2+q^2 \phi  $ &  \\ \hline
 7 & $X_1 \phi ^2+M_2 q \phi  \tilde{q}+M_1 \phi  \tilde{q}^2+M_1 M_2+M_2 M_3+q^2 \phi $ & \\ \hline
 8 & $X_1 \phi ^2+M_2 q \phi  \tilde{q}+M_1 \phi  \tilde{q}^2+M_4 q \tilde{q}+M_1 M_2+M_2 M_3+q^2 \phi  $ &  \\ \hline
 9 & $X_1 \phi ^2+M_1 \phi  \tilde{q}^2+M_3 q \phi  \tilde{q}+M_1 q \tilde{q}+X_2 \phi  \tilde{q}^2+M_2 X_3+M_3^2+M_1 M_2$  & \\ \hline
 10 & $X_1 \phi ^2+M_1 \phi  \tilde{q}^2+M_2 q \phi  \tilde{q}+q^2 \tilde{q}^2+M_1 M_2+M_2 M_3 $ &  \\ \hline
 11 & $X_1 \phi ^2 +M_1 q \phi  \tilde{q}+q X_2 \tilde{q}+M_1^2$  &  \\ \hline
 12 & $X_1 \phi ^2+M_1 \phi  \tilde{q}^2+M_2 q \phi  \tilde{q}+M_3 q \tilde{q}+M_1 M_2+M_2 M_3+M_2 M_4 $ &  \\ \hline
 13 & $X_1 \phi ^2+M_1 \phi  \tilde{q}^2+M_2 q \phi  \tilde{q}+q X_2 \tilde{q}+M_2^2 $ &  \\ \hline
 14 & $X_1 \phi ^2+M_2 q \phi  \tilde{q}+M_1 \phi  \tilde{q}^2+M_4 q \tilde{q}+q^3 \phi  \tilde{q}+M_1 M_2+M_2 M_3 $ &  \\ \hline
 15 & $X_1 \phi ^2+M_1 \phi  \tilde{q}^2+M_2 q \phi  \tilde{q}+M_3 q \tilde{q}+M_1 M_2+M_2 M_4 $ &  \\ \hline
 16 & $X_1 \phi ^2+M_1 \phi  \tilde{q}^2+M_2 q \phi  \tilde{q}+M_3 \phi  \tilde{q}^2+M_1 q \tilde{q}+M_1 M_2$ &  \\ \hline
 17 & $X_1 \phi ^2+ M_2 q \phi  \tilde{q}+M_1 \phi  \tilde{q}^2+M_4 q \tilde{q}+q^3 \phi  \tilde{q}+M_1 M_2+M_2 M_3+M_1 M_5$ & \\ \hline
 18 & $X_1 \phi ^2 +M_1 \phi  \tilde{q}^2+M_2 q \phi  \tilde{q}+M_3 q \tilde{q}+M_1 M_2$ &  \\ \hline
 19 & $X_1 \phi ^2+M_2 q \phi  \tilde{q}+M_1 \phi  \tilde{q}^2+M_4 q \tilde{q}+M_3 q^2 \phi +M_2 M_3+M_1 M_4+M_1 M_5+M_3 M_6$ &  \\ \hline
 20 & $X_1 \phi ^2+M_1 \phi  \tilde{q}^2+M_2 q \phi  \tilde{q}+M_3 q \tilde{q}+M_1 M_2+M_1 M_4$ &  \\ \hline
 21 & $X_1 \phi ^2+M_2 q \phi  \tilde{q}+M_1 \phi  \tilde{q}^2+q^3 \phi  \tilde{q}+M_1 M_2+M_2 M_3$ &  \\ \hline
 22 & $X_1 \phi ^2+M_1 \phi  \tilde{q}^2+M_2 q \phi  \tilde{q}+M_1 M_2+M_2 M_3$ & \\ \hline
 23 & $X_1 \phi ^2+M_2 q \phi  \tilde{q}+M_1 \phi  \tilde{q}^2+M_4 q \tilde{q}+M_5 q^2 \phi +M_1 M_2+M_2 M_3$ &  \\ \hline
 24 & $X_1 \phi ^2 +M_2 q \phi  \tilde{q}+M_1 \phi  \tilde{q}^2+M_3 q \tilde{q}+M_4 q^2 \phi +M_1 M_2$ & \\ \hline
 25 & $X_1 \phi ^2+M_1 \phi  \tilde{q}^2+M_2 q \phi  \tilde{q}+M_3 q \tilde{q}+M_1^2+M_2 M_3+M_2 M_4$ & \\ \hline
 26 & $X_1 \phi ^2+M_2 q \phi  \tilde{q}+M_1 \phi  \tilde{q}^2+q^3 \phi  \tilde{q}+M_4 q^2 \phi +M_1 M_2+M_2 M_3$ &  \\ \hline
 27 & $X_1 \phi ^2+M_2 q \phi  \tilde{q}+M_1 \phi  \tilde{q}^2+M_4 q^2 \phi +M_1 M_2+M_2 M_3$ & \\ \hline
 28 & $X_1 \phi ^2+M_1 \phi  \tilde{q}^2+q^2 \tilde{q}^2+M_1^2$ & \\ \hline
 29 & $X_1 \phi ^2+M_1 \phi  \tilde{q}^2+M_2 q \tilde{q}+M_1^2$ & \\ \hline
 30 & $X_1 \phi ^2+M_1 \phi  \tilde{q}^2+M_1 q \tilde{q}+M_1 M_2$ & \\ \hline
 31 & $X_1 \phi ^2+M_1 \phi  \tilde{q}^2+M_1^2$ & \\ \hline
 32 & $X_1 \phi ^2+M_2 q \phi  \tilde{q}+M_1 \phi  \tilde{q}^2+M_3 q \tilde{q}+M_4 q^2 \phi +M_1^2+M_2 M_3+M_2 M_5$ &  \\ \hline
 33 & $X_1 \phi ^2+M_1 \phi  \tilde{q}^2+M_1 q \tilde{q} $ &  \\ \hline
    34 & $X_1\tr \phi^2 +  M_1 \phi  \tilde{q}^2+ M_2 q \tilde{q} + M_3 \phi q\tilde{q} +M_1^2 $ & \\ \hline
 35 & $X_1 \phi ^2+M_1 \phi  \tilde{q}^2+M_2 \phi  \tilde{q}^2+M_1 q \tilde{q}$ & \\ \hline
 36 & $X_1 \phi ^2+M_2 q \phi  \tilde{q}+M_1 \phi  \tilde{q}^2+M_4 q \tilde{q}+M_3 q^2 \phi +M_3 X_2+M_2+M_2 M_3+X_1 X_3$ &  \\ 
    \hline 

  \caption{Corresponding superpotentials to Table \ref{tab:bad1}.} 
    	\label{tab:bad2}
      \end{longtable}
\end{center}

\end{appendix}

\end{document}